\documentclass[fleqn,usenatbib]{mnras}

\usepackage{newtxtext,newtxmath}

\usepackage[T1]{fontenc}

\DeclareRobustCommand{\VAN}[3]{#2}
\let\VANthebibliography\thebibliography
\def\thebibliography{\DeclareRobustCommand{\VAN}[3]{##3}\VANthebibliography}

\usepackage{graphicx}	
\usepackage{amsmath}	
\usepackage{color}
\usepackage[dvipsnames]{xcolor}
\usepackage{hyperref}

\newcommand{\hMpc}{\ \text{h}^{3} \ \text{Mpc}^{-3}}

\newcommand{\kmsmpc}{km\,s$^{-1}$Mpc$^{-1}$}

\title[Systematic Redshift Biases in BAO Cosmology]{The Effect of Systematic Redshift Biases in BAO Cosmology}

\author[Glanville, Howlett, and Davis.]{
Aaron Glanville,$^{1}$\thanks{E-mail: a.glanville@uq.edu.au}
Cullan Howlett,$^{1}$
Tamara M. Davis$^{1}$
\\
$^{1}$The University of Queensland, School of Mathematics and Physics, QLD 4072, Australia
}

\date{Accepted XXX. Received YYY; in original form ZZZ}

\pubyear{2020}

\begin{document}
\label{firstpage}
\pagerange{\pageref{firstpage}--\pageref{lastpage}}
\maketitle

\begin{abstract}
With the remarkable increase in scale and precision provided by upcoming galaxy redshift surveys, systematic errors that were previously negligible may become significant. In this paper, we explore the potential impact of low-magnitude systematic redshift offsets on measurements of the Baryon Acoustic Oscillation (BAO) feature, and the cosmological constraints recovered from such measurements. Using 500 mock galaxy redshift surveys as our baseline sample, we inject a series of systematic redshift biases (ranging from $\pm 0.2\%$ to $\pm 2\%$), and measure the resulting shift in the recovered isotropic BAO scale. When BAO measurements are combined with CMB constraints across a range of cosmological models, plausible systematics introduce a negligible offset on combined fits of $H_0$ and $\Omega_m$, and systematics must be an order of magnitude greater than this plausible baseline to introduce a $1 \sigma$ shift on such combined fits. We conclude that systematic redshift biases are very unlikely to bias constraints on parameters such as $H_0$ provided by BAO cosmology, either now or in the near future. We also detail a theoretical model that predicts the impact of uniform redshift systematics on $\alpha$, and show this model is in close alignment with the results of our mock survey analysis. 
\end{abstract}

\begin{keywords}
cosmology: cosmological parameters, distance scale, large-scale structure of Universe
\end{keywords}



\section{\label{sec:level1} Introduction}

Galaxy redshift surveys have provided a wealth of information on the nature of our universe. In particular, measurements of the two-point galaxy correlation function (and the associated galaxy power spectrum) have become a powerful tool to improve the strength of our cosmological constraints. As improved measurements and analysis techniques continue to reduce our statistical errors, it is more important than ever to ensure every link in this chain is robust to the increasingly significant impact of systematic errors. The need to tightly control systematics at every step in our analysis is particularly exemplified through the present tension between early and late universe constraints of $H_0$ \cite[currently at a $\sim 4-6 \sigma$ significance,][] {Riess2019Article}. This discrepancy has only become more entrenched with improved precision, spurring an effort to re-evaluate the potential impact of systematics that may have previously been neglected. Most of this effort has focused on potential systematics arising from standard candle {\em distance} calibrations, but recent work has also shown that surprisingly small offsets in {\em redshift} measurements ($\Delta z \sim 5 \times 10^{-4}$) can yield offsets of up to $1\%$ in the recovered value of $H_0$  \citep{Calcino2017, Davis2019}.  Small systematic biases in supernovae redshifts have also been found to significantly bias other cosmological parameters, such as $\Omega_m$ and $w$ \citep{Wojtak2015}. As such, it is worthwhile to review the accuracy of redshift measurements in Baryon Acoustic Oscillations, and how potential systematic errors in redshifts could contribute to cosmological tensions.

With the incredible precision provided by spectroscopic redshits, it can be easy to assume that redshift errors are negligible in surveys of large scale structure. Indeed, most redshifts are quoted to a precision of $10^{-3} - 10^{-4}$ in spectroscopic surveys, and are commonly provided without uncertainties. Whilst several BAO studies making use of photometric redshifts have carefully considered the impact of the significant error they can carry \citep{Marti2014, Sanchez2014, Beck2016}, being aware of large redshift errors and taking them into account (as in the case of photometric redshifts) is different to assuming redshifts are ideal and not quoting any errors (as is common in spectroscopic surveys).

In this work we focus on how potentially undiagnosed errors in redshift measurements themselves could bias constraints of the isotropic BAO scale, and the cosmological information recovered from such measurements. Our aim is to quantify how large any unresolved physical or observational effect must be to significantly bias galaxy clustering constraints, for comparison with known sources of redshift measurement error. In this way, our paper takes a similar approach to works such as \citet{Wojtak2015}, \citet{Calcino2017}, and \citet{Davis2019}, which detail the potential impact of systematic redshift errors on measurements of type Ia supernovae. Note that we will primarily focus on \textit{systematic} redshift errors (rather than the symmetric, Gaussian errors frequently quoted for individual redshifts) to distinguish errors that uniformly bias the apparent scale of the BAO feature (and by extension, uniformly bias our cosmological constraints), from errors that simply introduce additional uncertainty. 

Our paper is structured as follows. In Section 2, we begin with an overview of how measurements of standard rulers (in particular, the isotropic BAO feature) constrain cosmology. In Section 3, we detail every stage where redshifts are used in the measurement and analysis of the BAO feature, and review potential sources of systematic bias, along with their expected magnitude. We then use Sections 4 and 5 to explore how systematic redshift biases impact the measured scale of the BAO feature over a range of effective redshifts and galaxy number densities. Using the publicly available Multi-Dark Patchy mock surveys as our baseline sample, we inject biases (of varying magnitude) by systematically offsetting the mock catalogue redshifts. By extracting the BAO feature within samples containing an injected redshift bias, and comparing them to our baseline mock samples, we characterise how systematics affect the galaxy matter power spectrum and BAO scale. We then explore how these translate to offsets in cosmological constraints in Section 6, and assess whether plausible systematic biases introduce a large enough effect to merit consideration in current or future analyses.

\section{Background}

Standard rulers can broadly be described as objects with a known physical extent. Perhaps the most well known of these standard rulers, the BAO feature, corresponds to structures forming at a characteristic scale due to oscillations in the pre-recombination universe. These oscillations in the matter distribution of the early universe continued propagating until the drag epoch ($z \approx 1090$) where they were frozen at a fixed comoving scale. The overdensities associated with these oscillations supported a greater amount of galaxy formation than surrounding regions, creating an observable pattern in the distribution of galaxies today. This structure is observable as a clustering excess (`a bump') in the two point correlation function at a scale of $\sim 110 h^{-1}$ Mpc, or as a series of wiggles in the galaxy power spectrum. When decomposed into line-of-sight and perpendicular displacements, the anisotropic BAO feature corresponds to a standard ruler of length $\frac{c dz}{H(z)}$ and $\frac{D_A(z)}{d \theta}$ respectively, where $H(z)$ is the Hubble parameter, and $D_{A}(z)$ the angular diameter distance parameter. This $D_A$ parameter depends on curvature, encapsulated in $\Omega_k$, as
\begin{equation}
  D_{A} = (1+z)^{-1} 
    \begin{cases}
       R_0 \sinh (\chi) & \text{if $\Omega_k > 0$}\\
      R_0 \chi & \text{if $\Omega_k = 0$}\\
      R_0 \sin (\chi) & \text{if $\Omega_k < 0$}
    \end{cases}      \label{eq:DA} 
\end{equation}
where $\chi$ is the comoving distance, and $R_0=c/(H_0\sqrt{|\Omega_k|})$. It is important to note the distances used to construct such standard rulers are not measured directly, as redshift is the primary distance observable. Within the standard $\Lambda$CDM cosmological model, redshifts can be used to infer distances given some choice of fiducial cosmology,

\begin{equation}
    D(z) \equiv R_0\chi = \frac{c}{H_0} \int^z_0 \frac{dz}{E(z)}, \label{eq:Dz}
\end{equation}
\begin{equation}
    E(z) = \sqrt{\Omega_r (1+z)^4 + \Omega_m (1+z)^3 + \Omega_k (1+z)^2 + \Omega_{\Lambda}}. \label{eq:Ez}
    \end{equation}
    
In analyses where the line-of-sight and perpendicular displacements are not decomposed (the main focus of this paper), the standard length of the BAO feature ($D_V$) corresponds to the spherical average of these two decomposed standard distances,

\begin{equation}
    D_V = \left( (1+z)^2 \ D_A^2 (z) \ \frac{cz}{H(z)} \right)^{1/3}.
\end{equation}



Any difference between the fiducial cosmology used to infer this distance-redshift relationship and the true, underlying cosmology of our sample will lead to a dilation of the BAO feature. In an isotropic BAO analysis, this is encoded in $\alpha$, which is fit by dilating a fiducial template power spectrum or correlation function against our model. This best-fitting $\alpha$ then encodes the difference between our model and fiducial cosmologies as
\begin{equation}
    \alpha = \frac{D_V (z) r_s^{\rm{fid}}}{D_V^{\rm{fid}} (z) r_s}, \label{eq:alpha}
\end{equation}
(where $r_s$ is the sound horizon radius at the drag epoch). By fitting $\alpha$ (that is, finding the dilated template that best fits our model), one can quantify how far the underlying cosmology of a sample deviates from the chosen fiducial, and use this to derive the best-fitting sample cosmology.

\section{Redshifts and BAO Analysis}

Throughout the process of measuring and analysing the BAO feature, a variety of individual and `effective' redshifts are employed. We use this section to clarify where these redshifts are introduced, and detail some conventions in their application. We also describe some potential sources of error at each stage of analysis, their magnitude, and how these are expected to affect the BAO feature. 

\subsection{Observational Redshifts}

Measurements of the BAO feature begin with a galaxy redshift survey, a catalogue of the angular sky position and redshift of hundreds of thousands of individual sources. Utilizing a fiducial cosmology to define a distance-redshift relationship (as per Equations \ref{eq:Dz} and \ref{eq:Ez}), these points are used to populate a 3D Cartesian space. The full power spectrum or correlation function (including the imprinted BAO feature) is then measured from this projected 3D sample. At this observational stage, we broadly categorise redshift errors as either symmetric (i.e. errors without any preferred direction), or systematic (errors that uniformly affect the average redshift of a sample).

\subsubsection{Photometric/Spectroscopic Fitting}

Galaxy surveys commonly measure redshifts by either directly fitting the emission and absorption lines of a fully resolved spectrum (yielding a spectroscopic redshift), or by measuring the brightness of a source propagated through discrete passband filters (yielding a photometric redshift). Both techniques are subject to observational errors that can impact cosmological analyses.

The highest precision redshifts used in galaxy redshift surveys are provided by fitting the emission/absorption lines of a fully resolved spectra. Any mismatch in the spectral lines between the data and comparison template (i.e. the incorrect identification of spectral features) can result in a high-magnitude, catastrophic redshift error. The redshifting software used by the SDSS team (for both BOSS and eBOSS) was designed to recover a $<1\%$ catastrophic redshift error rate (where ``catastrophic failures'' correspond to unrecognised redshift errors $\ > \ 1000 \rm{km s^{-1}}$ for galaxies, or a $\sigma_z$ of $\approx 10^{-2} - 10^{-3}$) \citep{Bolton_2012, Hutchinson_2016}. Very recently, \citet{Massara2020} explored how one such catastrophic error (the misidentification of a H ${\beta}$ emission line as an O {\sc III} source) could introduce a notable bias in redshift measurements made using the Roman Space Telescope. In particular, \citet{Massara2020} highlighted the unique significance of such a catastrophic redshift failure in the context of BAO analyses, with this misidentifcation offsetting the inferred distance to a source by $\sim 90 h^{-1} \text{Mpc}$, systematically biasing the best-fitting scale of the BAO feature. Using an estimated H ${\beta}$ -- O {\sc III} interloper fraction of $4-5\%$, \citet{Massara2020} determine that such catastrophic redshift errors can introduce systematic biases on $\alpha$ of $\sim 1\%$ in the $1.3 < z < 1.9$ redshift range for a Roman-like survey. While H ${\beta}$ -- O {\sc III} interlopers may be present in future Roman-like surveys, fits on spectral lines, and measurements with a higher signal-to-noise ratio (which can more reliably resolve the O {\sc III} double peak) will significantly reduce this interloper fraction. 

While mismatched spectral lines provide the highest magnitude errors, even samples where the spectral features have 
been correctly identified contain a small degree of uncertainty in the redshift fit, arising due to scatter in redshift fitting, photon noise, and spectral resolution limitations. \citet{Bolton_2012} find that statistical uncertainties arising from effects such as photon noise are on the order of a few tens of $\rm{km s^{-1}}$ in the BOSS sample (corresponding to a $\sigma_z$ on the order of $~10^{-4}$), with similar results for the eBOSS sample provided in \citet{Hutchinson_2016}. Additionally, since most spectrographs typically have a redshift resolution of $10^{-3} - 10^{-4}$, redshift scatters and errors below these magnitudes are likely present in galaxy redshift data.

In general, catastrophic redshift mismatches and intrinsic measurement scatters (for both photometric and spectroscopic redshifts) have no average preferred direction. Such symmetric errors primarily work to dampen the strength of the BAO feature, increasing the uncertainty of our cosmological constraints. However these do not uniformly shift the position of this feature or, by extension, the best-fitting cosmology. While it is possible for catastrophic redshift failures to exhibit a slight directional bias (for example, due to non-negative priors on redshift), most of these failures are uncorrelated with the scale of the BAO feature. Cases where such catastrophic failures are potentially correlated with the BAO scale \citep[such as the H ${\beta}$ -- O {\sc III}  mismatch detailed in][]{Massara2020} are mitigated by fitting on multiple spectral lines and high signal-to-noise ratios. While these effects represent an important source of observational error, we do not further explore the effect of these broadly symmetric effects in our paper. 

\subsubsection{Comoving Frame Corrections}

Redshifts quoted in galaxy redshift surveys are generally quoted relative to the rest frame of the CMB, and so must be converted from the heliocentric rest frame (commonly provided at the instrument level). In order to convert from a heliocentric to CMB rest frame, one must remove the redshift arising from the peculiar velocity of the Sun with respect to the CMB rest frame from the observed redshift,

\begin{equation}
    z_{\rm{CMB}} = \frac{1 + z_{\rm{obs}}}{1 + z^{\rm{pec}}_{\rm{sun}}} - 1 \label{eq:HelioCorrection}
\end{equation}

\begin{figure}
    \centering
        \includegraphics[width=8.6cm]{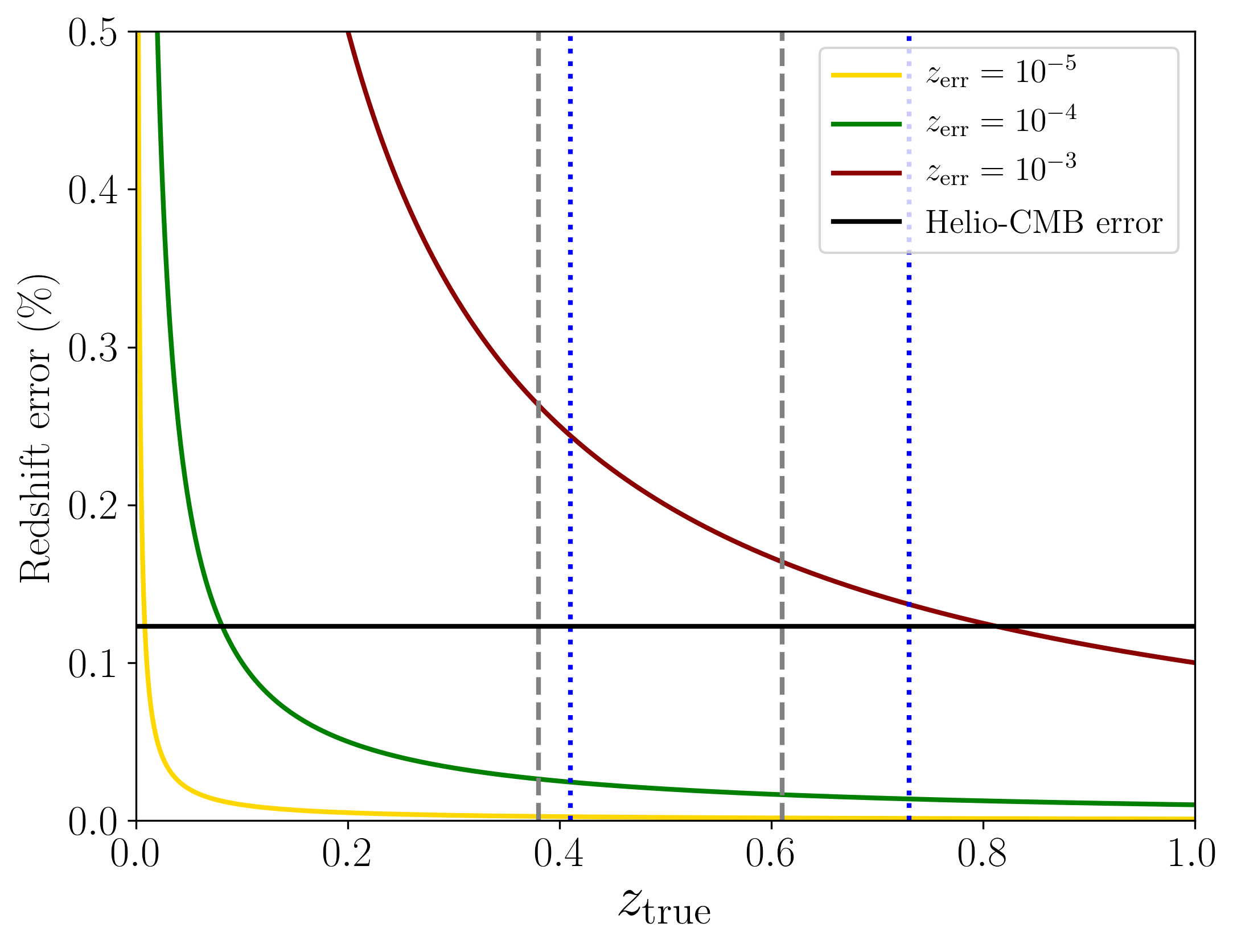}
        \caption[]{Impact of various redshift systematics as a function of their true redshift, intercepted by the maximum and minimum BAO $z_{\text{eff}}$ quoted by the BOSS (grey dashed) and WiggleZ (blue dotted) surveys. The use of a low-$z$ approximation in heliocentric-CMB reference frame corrections introduces a $>0.1\%$ redshift bias across all bins, corresponding to an absolute redshift offset of around $10^{-3}$-$10^{-4}$. Restrictions in spectral resolution mean we also consider the effect of potentially unnoticed effects of a magnitude $10^{-3}$-$10^{-4}$ in redshift. Over the effective redshift range probed by surveys such as BOSS and WiggleZ, we use this to place an upper limit on plausible systematics at an order of $0.2\% - 0.4\%$ for the purposes of this work.}
        \label{fig:RedshiftErrorpc}
\end{figure}

The redshifts recovered from this conversion process can be systematically biased if the low redshift approximation ($z_{\rm{CMB}} \approx z_{\rm{obs}} - z^{\rm{pec}}_{\rm{sun}}$) is mistakenly applied in place of the full redshift addition formula, introducing an offset on the order of $\Delta z \approx 10^{-3} - 10^{-4}$, depicted in Figure \ref{fig:RedshiftErrorpc} \citep{Davis2019}. The use of this low redshift approximation in sources such as the NED velocity conversion calculator make it likely that this systematic is present in the literature, and so we consider observational level redshift offsets of this magnitude plausible throughout our analysis (\citealt{Davis2019}; Carr et al. in prep.) 



\subsubsection{Physical Effects}

Beyond simple observational errors and misapplied corrections, a number of physical effects can systematically offset our measured redshifts. One possible physical effect that has emerged in the literature is the potential impact of living in a gravitational over/underdensity, and the associated blue/redshift of light measured in such a potential. \citep{Wojtak2015, Calcino2017}. The total effect of these local gravitational shifts can be characterised as the sum of the gravitational redshift when light exits a source, and the gravitational blueshift acquired when light enters our local potential well. As such, living in a statistical overdensity will, on average, result in a slight blueshifting of all observed sample. Similarly, living in a local underdensity will result in a systematic additional redshift. Cosmological $N$-body simulations indicate such local gravitational effects could systematically offset observed redshifts by $\sim 10^{-5}$, an order of magnitude smaller than the observational effects we discussed previously \citep{Wojtak2015}. In addition to the configuration of our local universe, high-velocity ($>1000 \text{km s}^{-1}$, \citealt{Trump2006}, \citealt{Gibson2009}) Quasar outflows along the line of sight direction could work to systematically bias high-$z$ measurements of the BAO feature. While such high-$z$ measurements are infrequently used to explicitly constrain parameters such as $H_0$, these effects may be worth considering as future surveys probes high-$z$ structure at greater precision. As such, the relative impact of such physical effects are likely to be negligible in comparison to redshift offsets introduced by observational/correction effects. 


\subsubsection{Unknown or Unresolved Effects}

In summary, the most significant source of systematic redshift bias we consider is the use of a low redshift approximation when correcting to the CMB rest frame from the heliocentric frame (introducing a potential offset of $\Delta z \approx 10^{-3} - 10^{-4}$). While this is the most significant known source of bias we consider, spectroscopic redshifts in galaxy surveys are restricted by their spectral resolution (at an order of ~$10^{-3} - 10^{-4}$). For example, the BOSS team quote their spectral resolution ($R$) to $1500-2600$ (corresponding to a precision in redshift of $6.7 \times 10^{-4} - 4 \times 10^{-4}$), whereas the WiggleZ spectra are measured at a resolution of $R = 1300$ ($\sigma_z \approx 7.7 \times 10^{-4}$, \citealt{Drinkwater_2010}), and the upcoming DESI survey anticipates spectral resolutions ranging from $2000 - 5500$ ($\sigma_z \approx 5 \times 10^{-4} - 1.8 \times 10^{-4}$, \citealt{DESI_2016}). As such, physical or observational effects that systematically offset redshifts at an order below this spectral resolution could plausibly go unnoticed. This means that in addition to the various known sources of potential systematic redshift bias, unresolved effects could induce an offset of up to $\Delta z \approx 5 \times 10^{-4}$, which we consider moving forward in our analysis. 

\subsection{Random Catalogues}

Measurements of galaxy clustering require random, unclustered catalogues to estimate the expected number of galaxies at the location of each real galaxy. Furthermore, when we measure the clustering of galaxies in a redshift survey, this signal includes a number of non-physical effects (arising from our survey geometry, the discrete gridding of the survey volume etc.). In order to account for these effects, we typically convolve our model power spectrum with a window function, constructed using a random catalogue distributed over the same volume and sky patch as our survey. Since this random catalogue must accurately approximate the angular and redshift distributions of the real dataset, some care must go into the specific formalism used to assign redshifts to our random catalogue, and the impact of any choice on our measurements and our model power spectrum after convolution. 

Broadly, the redshifts of sources in random catalogues are assigned by either ``shuffling'' though observed redshifts, or by sampling from a smoothed spline fit to the observed galaxy distribution. In the shuffling method, a random point is picked within the observational sky patch (i.e. the angular sky co-ordinates are chosen at random), and then a redshift is assigned to a ``source'' at this point by randomly selecting a true redshift from the observed sample. In the spline method, the observational data is binned by redshift, and an N-node spline fit is used to roughly approximate the underlying $n(z)$. The redshift for each random angular sky point is then assigned by sampling from this spline fit, after smoothing. \citet{Ross_2012} tested the impact of these choices on the correlation function recovered from mock galaxy catalogues, and found a low-level systematic bias and uncertainty is introduced through the use of algorithmically assigned random redshifts (where offsets were more pronounced on the quadrupole fit than the monopole fit). They also found that the shuffle method provides the smallest systematic bias and uncertainty. Since the impact of these choices in redshift assignment are understood, we do not explicitly consider their impact in this paper. However, given that observational data is used to construct this $\bar{n}(z)$ profile, any redshift bias that affects this number density distribution will also be present in the random catalogue. As such, when biasing any mock catalogues in our paper, we also apply this effect to our random catalogue when making clustering measurements and when creating any window functions.

\subsection{Redshift of the BAO Feature}

After measuring the galaxy power spectrum of our observational data, the next step to fitting $\alpha$ is to extract the BAO feature from this signal. Once the scale of this feature is measured from the extracted BAO signal, it is compared to the expected scale of the fiducial cosmology (as per Equation \ref{eq:alpha}). 

In most models, the BAO feature is extracted from the galaxy power spectrum through comparison with a smoothed linear spectrum evaluated at the effective redshift of the data $z_{\text{eff}}$. A number of different conventions exist for the definition of the $z_{\text{eff}}$, however in general it corresponds to the weighted average of redshifts within a given sample,

\begin{equation}
    z_{\rm{eff}} = \frac{\Sigma_i \bar{z}_i w_i}{\Sigma_i w_i}.
\end{equation}
In detail, different surveys make slightly different choices in the redshifts they use to evaluate $z_{\text{eff}}$. For example, the WiggleZ team \cite{Blake_2011} use the mean redshift of galaxy pairs to calculate $\bar{z}$,

\begin{equation}
    \bar{z}_i = \frac{(z_{1,i} + z_{2,i})}{2},
\end{equation}
whereas the BOSS team use the average of individual galaxy redshifts \citep{Beutler_2016}. Further, methods of evaluating $z_\text{eff}$ typically neglect the evolution of $D_A(z)$ and $H(z)$ across the redshift range of the sample, assuming that the chosen bins are small enough to make any redshift dependence negligible. In general however, robust fits on the BAO scale do not rely on a precise evaluation of $z_{\text{eff}}$.

Broadly, this insensitivity arises as a result of the fact that $\alpha$ is defined as the ratio of the recovered BAO scale in the test model compared to the expected scale in some fiducial model. Any difference in the convention used to define $z_{\text{eff}}$ will apply to both the model, and fiducial scales. In short, comparing these standard rulers at different scales will not strongly affect our final results, since we are primarily interested in the $\textit{ratio}$ of our model and fiducial scales. As such $z_{\text{eff}}$ is frequently only quoted to two decimal places, and without uncertainties. 

\begin{figure*}
    \centering
        \includegraphics[width=8.0cm]{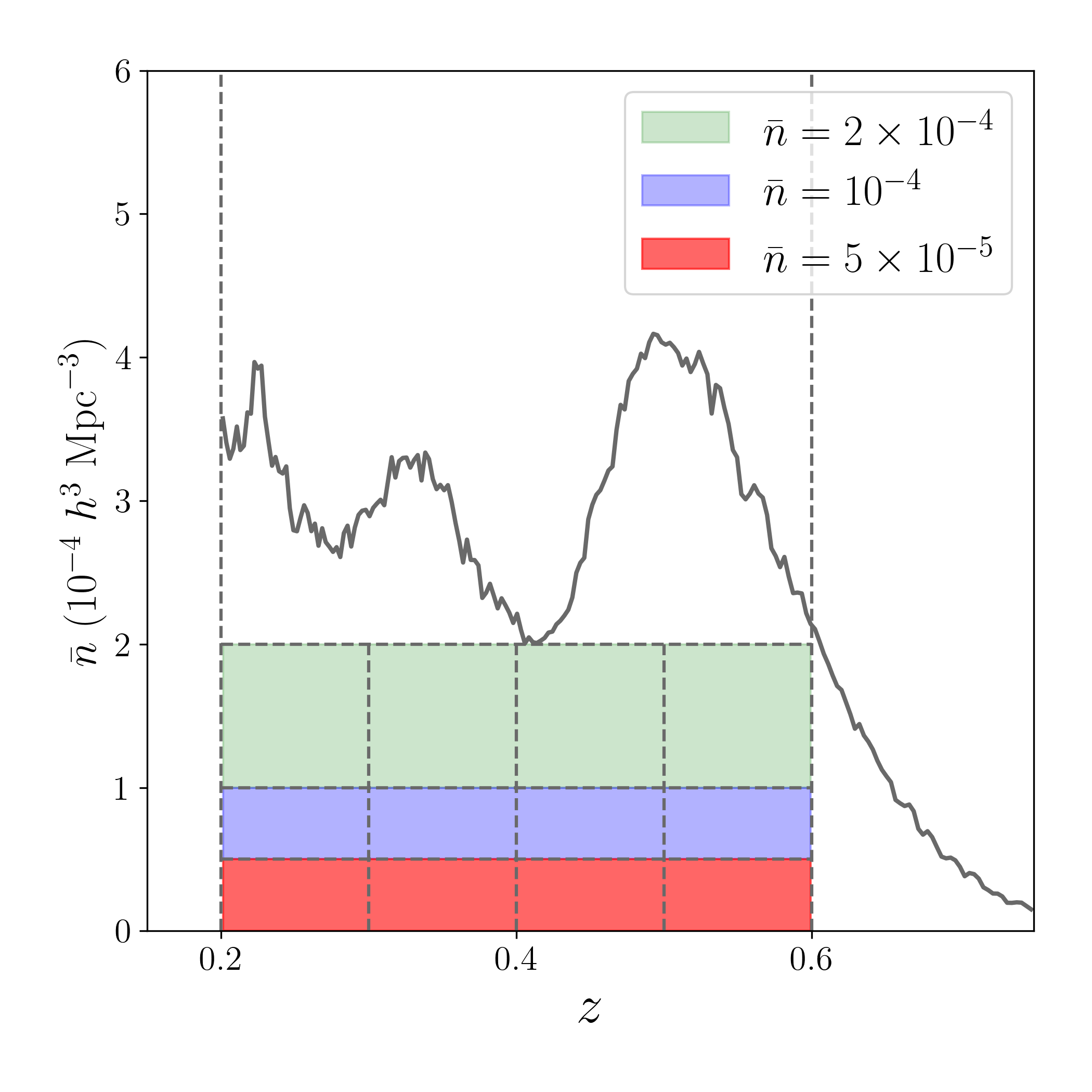}
        \includegraphics[width=8.0cm, height=8.0cm]{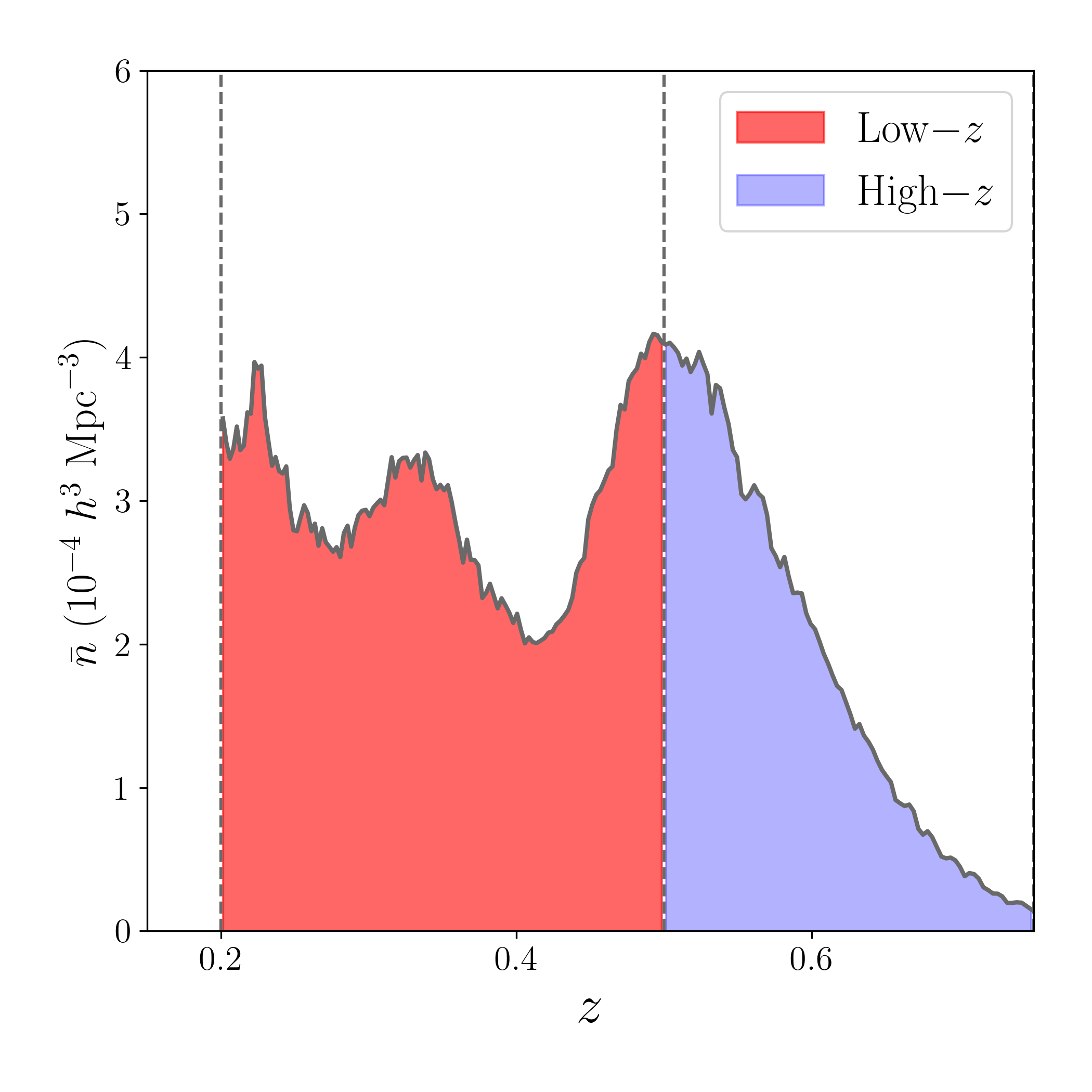}
        \caption[Survey/Mock Number Density]{Number density ($\bar{n}$) as a function of redshift for one of the 500 MultiDark Patchy mock data sets used in this analysis. Throughout our analysis, we broadly apply two different data cuts to these mocks--- In order to explore the significance of $\bar{n}$ on the sensitivity of mocks to injected bias (as in Figure \ref{fig:BiasedAlpha}), we subsample our data to create catalogues of constant $\bar{n}$ as a function of redshift (left), ranging from $0.2<z<0.6$. When studying the effect of injected systematics on BOSS-like samples (as in Figure \ref{fig:NGCAlphaOffset}), we bin the entire mock catalogue by redshift (right), generating a low-$z$ ($0.2<z<0.5$) and high-$z$ ($0.5<z<0.75$) sample.}
        \label{fig:NumDensityComparison}
\end{figure*}

\subsection{Summary}

Both individual and averaged redshifts are introduced at a number of different stages within the measurement and analysis of the BAO feature--- as such, there are a number of potential vectors where redshift biases can be introduced into our analysis pipeline. Broadly however, we are most interested in the potential impact of observational level systematics at an order of $z_{\text{err}} = 10^{-3} - 10^{-4}$, corresponding to an upper limit systematic $\sim 0.4\%$ in the redshift regime probed by standard BAO measurements (Figure \ref{fig:RedshiftErrorpc}). We note that due to the limitations of spectral resolution in most galaxy redshift surveys, additional systematic effects at this threshold could remain undetected in the data. 

\section{Methodology}

We have established an upper threshold of known redshift systematics, and the various points in a BAO analysis pipeline where these systematics could be introduced. We now focus on how such errors could bias measurements of the isotropic BAO feature, beginning by constraining the total minimum systematic error required to significantly impact cosmology. We then review whether such a total bias could plausibly arise through a combination of known systematics, and effects that could be potentially unresolved due to spectral limitation. 

\subsection{Galaxy Redshift Catalogue}

In order to explore the impact of redshift systematics arising at the observational/instrumentation stage, we begin with a simulated galaxy redshift catalogue. For this purpose, we make use of the publicly available MultiDark-Patchy mock datasets, created for the $12^{\text{th}}$ SDSS data release \citep{Kituara2016, RodriguezTorres2016}. These mock catalogues are designed to replicate the redshift range ($0.2<z<0.7$), number density, and survey geometry of the galaxy redshift catalogue used by the BOSS team \citep{DESI_2016}. The Patchy mock catalogues used in this analysis are produced using a fixed cosmology ($\Omega_m$ = 0.307115, $\Omega_{\Lambda}$ = 0.692885, $\Omega_b$ = 0.048, $\sigma_8$ = 0.8288, $h$ = 0.6777), and include a variety of physical and observational effects (such as redshift space distortions, fiber collisions, and the evolution of clustering with redshift). The BOSS results remain some of the most precise BAO measurements to date. As such, the MultiDark-Patchy mocks provide an optimal mock observational survey to test the effects of systematic redshift biases as they are propagated throughout a standard analysis pipeline. The Patchy mock datasets also provide random catalogues (that replicate the survey geometry and number density of this mock data), with redshifts assigned using the shuffle method to minimise systematic error. In our work, we use 500 Patchy mock surveys as our baseline sample. We apply cuts to this data (as per Figure \ref{fig:NumDensityComparison}) and offset the measured redshifts of these samples using either a multiplicative offset ($z_\text{bias} = z \times (1 \pm \text{bias})$) or a constant offset ($z_\text{bias} = z \pm \text{bias}$). This allows us to simulate a range of systematic effects (varying in both magnitude and type), for direct comparison with our unbiased baseline catalogues.

\begin{figure}
    \centering
        \includegraphics[width=8.0cm]{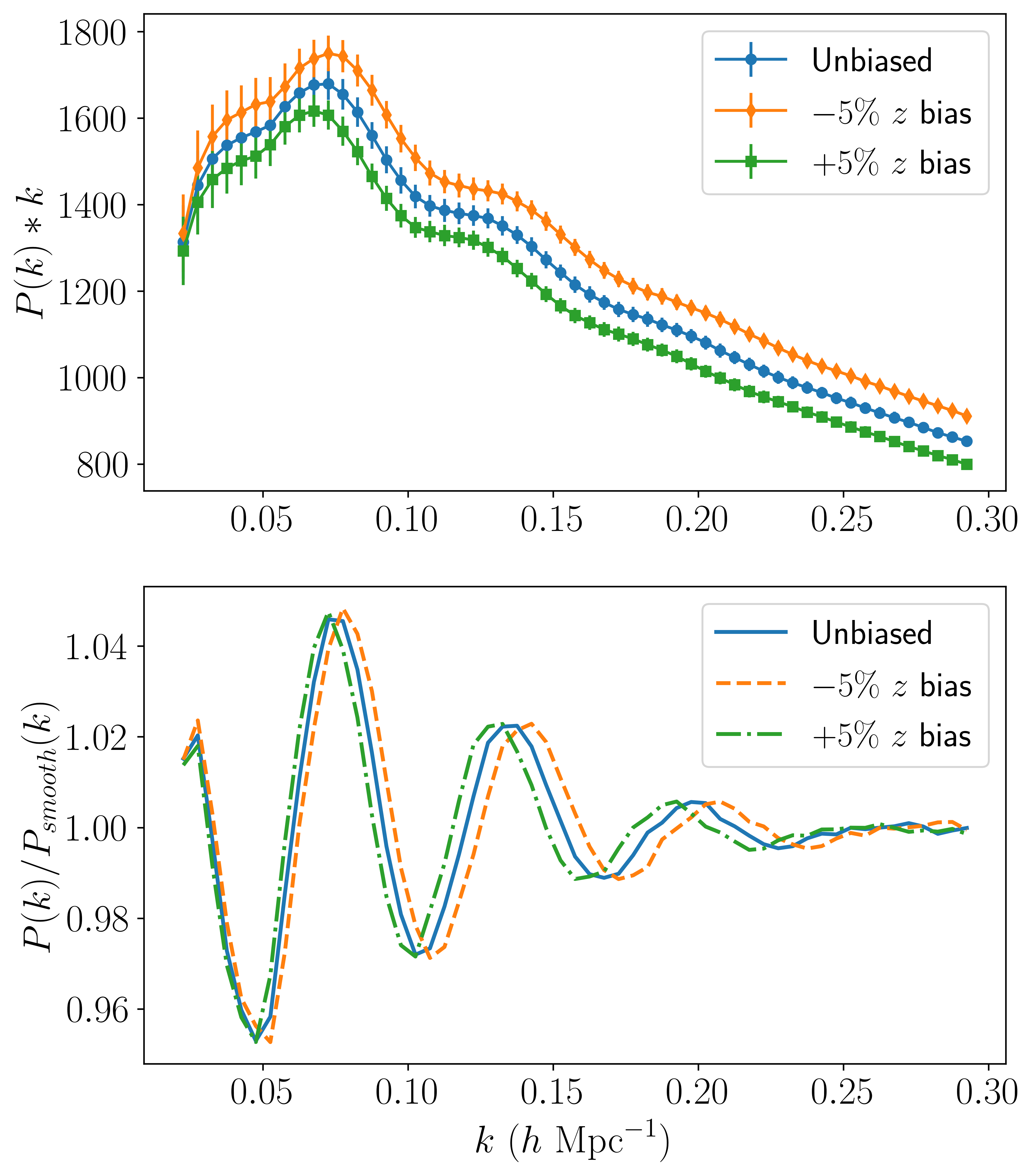}
        \caption[]{Average of 500 galaxy power spectra evaluated using default Patchy mock data and extremely redshift-biased samples (upper), and the extracted BAO features for each population (lower). This plot highlights two key features from a systematic redshift bias on the 1D power spectrum--- a shift in the overall amplitude of the power spectrum (up or down) and a dilation of the extracted BAO features along $k$ (left and right). Uniformly decreasing the redshifts of our samples pushes the peaks of our power spectrum to smaller scales (higher $k$ modes, lower $\alpha$), whereas uniformly increasing our source redshifts pushes our peaks to larger scales (lower $k$ modes, higher $\alpha$).}
        \label{fig:BiasedWiggles}
\end{figure}

\begin{figure*}
    \centering
        \includegraphics[width=8.0cm]{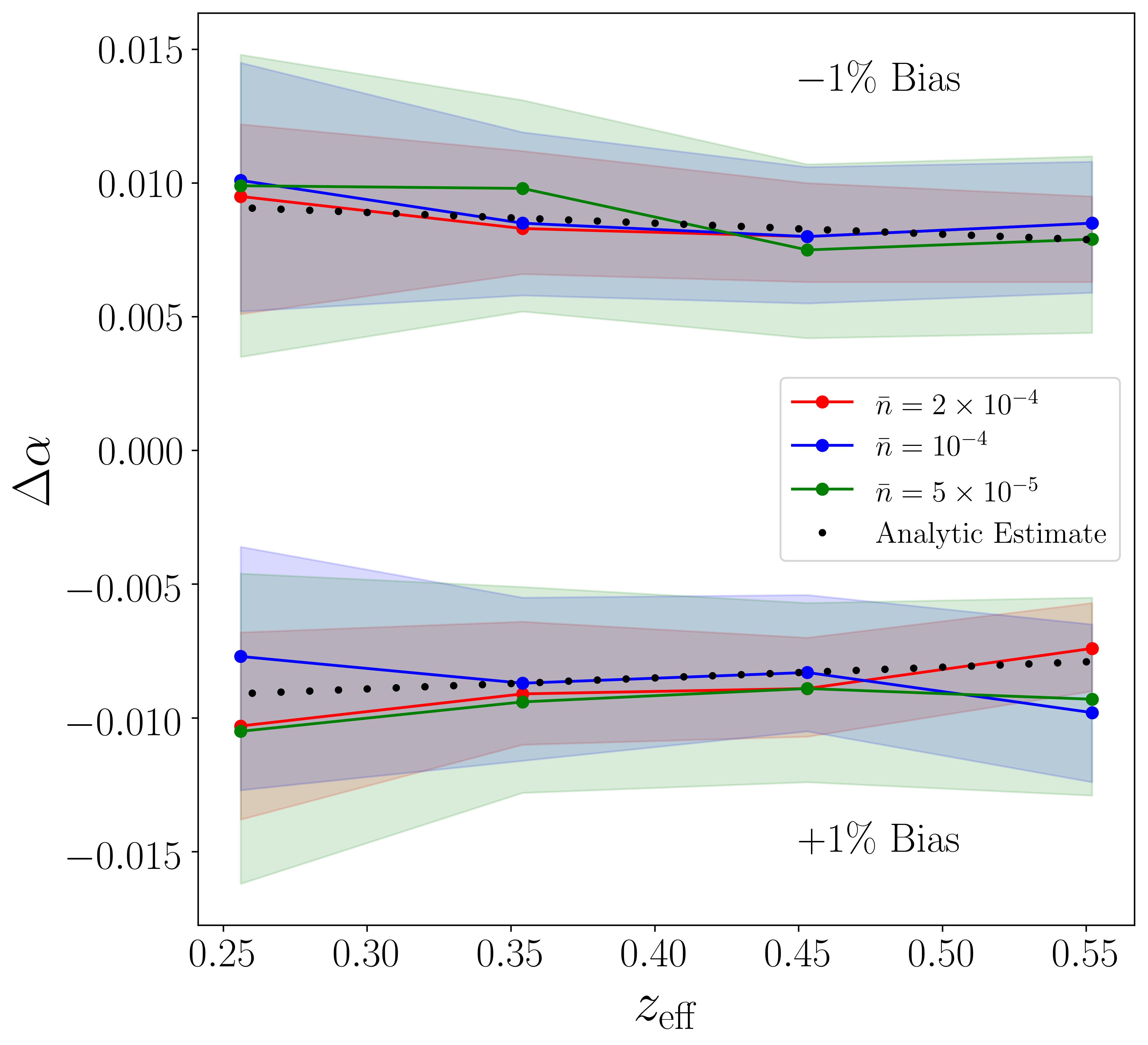}
        \includegraphics[width=8.0cm]{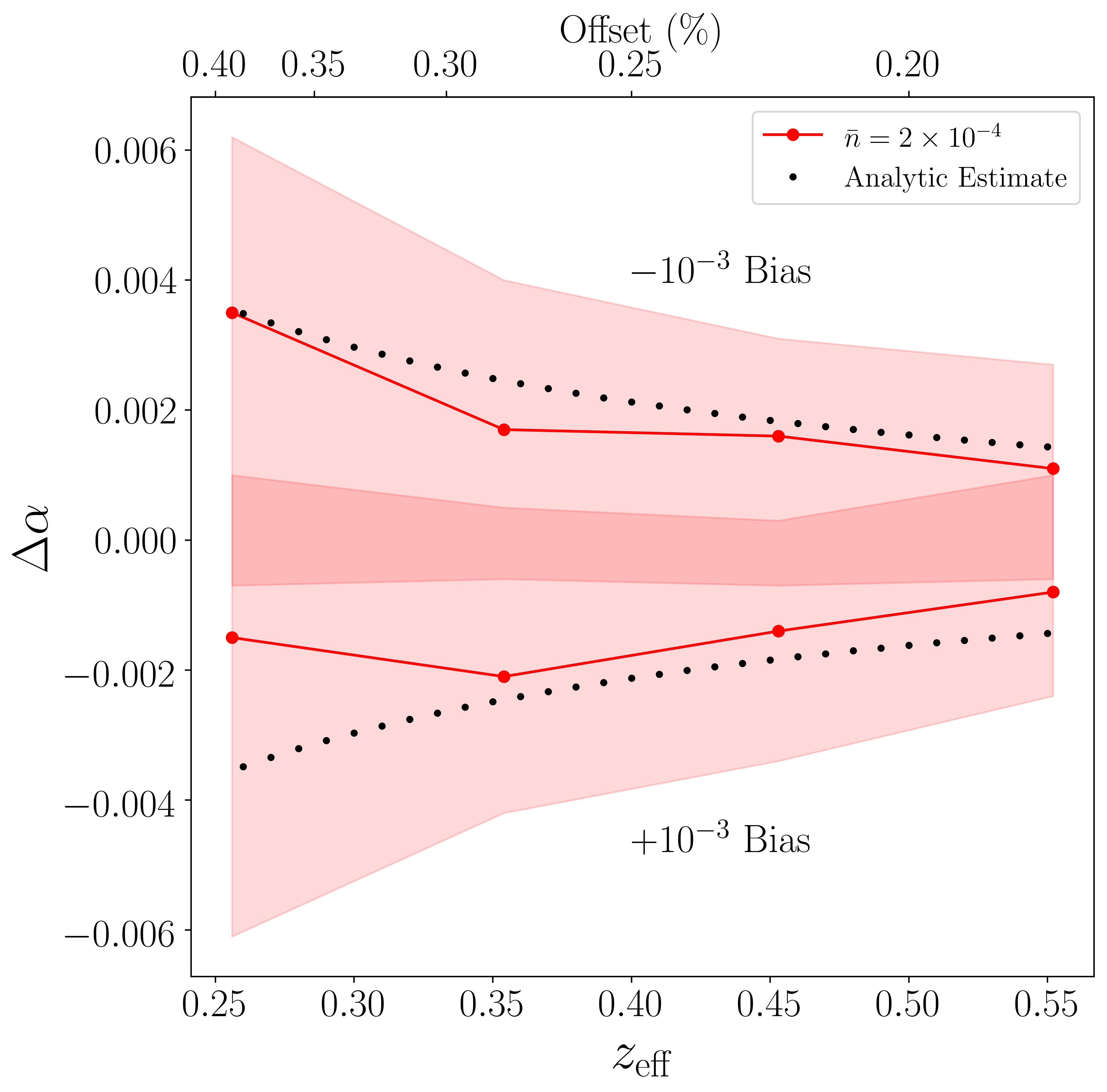}
        \caption[]{The effect of a $\pm 1\%$ observational level redshift offset (left) and $\pm 10^{-3}$ bias (right) on the average $\alpha$ of 500 mock surveys as a function of $z_{\text{eff}}$. Note that the mock data used in this analysis was subsampled to yield a flat number density across all redshift bins (with tested densities ranging from $\bar{n} = 2 \times 10^{-4} \hMpc$ to $5 \times 10^{-5} \hMpc$). The injection of a $\pm 1\%$ multiplicative offset in observed redshifts results in an almost uniform bias in $\alpha$ from our reference samples, across the range of tested $z_{\text{eff}}$ and $\bar{n}$. This is in good agreement with our analytic model for such errors (detailed in Appendix A, Equation \ref{eq:AnalyticAlphaApplied}), and confirms $\bar{n}$ has no appreciable impact on the magnitude of this bias. As such, we make use of the highest number density sample ($\bar{n} = 2 \times 10^{-4} \hMpc$ moving forward. When we inject a fixed, constant offset of $10^{-3}$ to the redshifts in this high $\bar{n}$ sample, we see a larger offset in $\alpha$ from our baseline sample for lower $z_{\text{eff}}$ (corresponding to the increase in bias as a proportion of total measurement with lower $z_{\text{eff}}$), in good agreement with our analytic model.}
        \label{fig:BiasedAlpha}
\end{figure*}

\subsection{From Survey Data to $\alpha$}

We project the point sources from these galaxy redshift catalogues into a 3D box and measure the power spectrum of this sample using the publicly available \textsc{Nbodykit} code suite \citep{Hand2018}. When populating a Cartesian box from this mock observational data, we use the underlying cosmology of the Patchy mock samples (i.e. the cosmology used to construct these mocks) to define our distance-redshift relationship, and weight the resulting points using the FKP scheme \citep{FKP1994}. Note that after biasing mock catalogues, we do not explicitly recompute the number density of these samples. The change in catalogue $\bar{n}(z)$ arising from redshift biasing may slightly change the magnitude of assigned FKP weights if updated, however since this effect is broadly uniform across all weights it was considered a marginal effect. 

In order to extract the BAO feature from the matter power spectra of our samples, we make use of the publicly available BAO fitting code \textsc{Barry} \citep{Hinton2020}. \textsc{Barry} provides a modular framework for fitting the BAO features in the galaxy power spectrum and correlation function, allowing us to easily compare the effects of different redshift ranges, datasets, and BAO extraction models from the literature. The effect of a multiplicative redshift offset on the average of our 500 Patchy mock datasets (and the BAO feature extracted through \textsc{Barry}) is illustrated in Figure \ref{fig:BiasedWiggles}.


By applying some uniform offset in redshift to our mock catalogue data, the underlying distances and scales of our sample are represented at a different apparent redshift. In effect, we are constructing an apparent distance-redshift relationship that deviates from the true distance-redshift relationship of our sample. Since this biasing is rooted in this apparent distance-redshift relationship, it is possible to analytically evaluate the expected impact of our biasing scheme on $\alpha$, for comparison with fits provided from our mock catalogues. Broadly, we can model this change in the apparent distance-redshift relationship of our sample as a modification of Equation \ref{eq:alpha},

\begin{equation}
    \alpha_{\text{bias}}(z+\Delta z) = \dfrac{D_V(z) r_s^{\text{fid}}}{D_V^{\text{fid}}(z + \Delta z) r_s}. \label{eq:alphabiastwo}
\end{equation}
In the limit of low redshifts, this bias function can be roughly approximated as $\Delta \alpha = \alpha_{\text{bias}} - \alpha_{\text{base}} \approx \frac{- \Delta z}{z}\alpha_{\text{base}}$ (we detail the full analytic evaluation of this bias function in Appendix A).

Our technique of injecting a uniform bias in observed redshifts shares some similarities with the galaxy redshift catalogue blinding scheme put forward by \citet{Brieden2020}. In this scheme, galaxy redshifts are blinded by first converting observed redshifts to distances using some bias cosmology ($\Omega^{\text{bias}}$), before inverting these distances back to redshifts using a second, reference cosmology ($\Omega^{\text{ref}}$). This blinding technique results in a homogeneous dilation in blinded redshifts from the original catalogue (seen in Figure 4 of their paper), similar to the output of our direct redshift biasing technique. A key concept of their blinding scheme is that the bias in $\alpha$ can then be removed by multiplying by the appropriate factors of $D_{V}^{\rm{bias}}(z)$ or $D_{V}^{\rm{ref}}(z)$. From Equation \ref{eq:alphabiastwo} one can see that if we were to multiply the biased $\alpha$ value by $D_{V}^{\rm{fid}}(z+\Delta z)$ we too would remove any bias. When using BAO measurements for cosmological constraints, this \textit{is} actually what we do and so one would naively think that the bias in $\alpha$ does not propagate through into cosmological constraints. Unfortunately, the bias reappears because we then go on to assume we have measured a distance to the biased redshift $D_{V}(z+\Delta z)$, whereas from Equation~\ref{eq:alphabiastwo} it is clear we have in fact measured $D_{V}(z)$ (see also Section~\ref{sec:cosmo}).

\section{Results}

\subsection{Effect of Systematic $z$ Bias on $\alpha$}

Before analysing the impact of an injected redshift bias within our full Patchy mock sample, we must first determine the potential role that sample size and $z_{\text{eff}}$ plays in the sensitivity of $\alpha$ to redshift systematics. As such, we begin by studying the impact of injected observational redshift biases on galaxy surveys that vary in $z_{\text{eff}}$ and number density, $\bar{n}$. To generate catalogues that vary in ($z_{\text{eff}}, \ \bar{n}$), we begin by splitting our baseline sample of 500 Patchy mock surveys into redshift bins of width 0.1 (extending from $z = 0.2-0.3$ to $z=0.5-0.6$) as in Figure \ref{fig:NumDensityComparison}, left. We then further subsample from these binned datasets to generate galaxy redshift catalogues with a fixed number density as a function of redshift (ranging from $\bar{n} = 5 \times 10^{-5} \hMpc$ to $\bar{n} = 2 \times 10^{-4} \hMpc$). This provided a series of baseline galaxy surveys varying in number density and $z_{\text{eff}}$, that were individually measured, analysed, and used to fit for $\alpha$. 

With these subsampled mocks as our reference catalogues, we explored the effect of a multiplicative ($\pm 1\%$) and constant ($\pm 10^{-3}$) redshift systematic on the value of $\alpha$ recovered for each choice of $\bar{n}$ and $z_{\text{eff}}$. For each choice of $\bar{n}$ and $z_{\text{eff}}$, we subtract the $\alpha$ recovered from each biased sample from its corresponding baseline, $\Delta \alpha(\bar{n}, z_{\text{eff}}) = \alpha_{\text{bias}} (\bar{n}, z_{\text{eff}} + \Delta z) - \alpha_{\text{base}}(\bar{n}, z_{\text{eff}})$, and plot this as a function of $\bar{n}$ and $z_{\text{eff}}$ in Figure \ref{fig:BiasedAlpha}. We find that the average shift in $\alpha$ arising from an injected multiplicative bias (left, $z_{\text{bias}} = z \times (1 \pm 10^{-2})$) is broadly independent of the number density and effective redshift of our sample, as predicted in our analytic model. While $\bar{n}$ and $z_{\text{eff}}$ do not significantly impact the sensitivity of $\alpha$ to an injected bias, they do play a significant role on the uncertainty of $\alpha$. Since samples with a higher $\bar{n}$ and $z_{\text{eff}}$ have a greater number of sources over a larger cosmological volume, these catalogues have a reduced statistical error. We also find a good fit to our analytic model in tests with an injected bias that remains constant with redshift (right, $z_{\text{bias}} = z \pm 10^{-3}$). In this case, we see the offset in $\alpha$ from our reference sample decreases with increasing $z_{\text{eff}}$. This inverse curve arises due to the decreased fractional magnitude of our constant $10^{-3}$ offset as a function of $z_{\text{eff}}$, consistent with our prediction. These results suggest that isotropic BAO fits are generally sensitive to the magnitude of systematic biases as a function of their effective redshift ($\frac{\Delta z}{z_{\text{eff}}}$), which is consistent with our theoretical model for these errors developed in Appendix A. 

\subsection{BOSS Samples}

\begin{figure}
    \centering        
    \includegraphics[width=8.0cm]{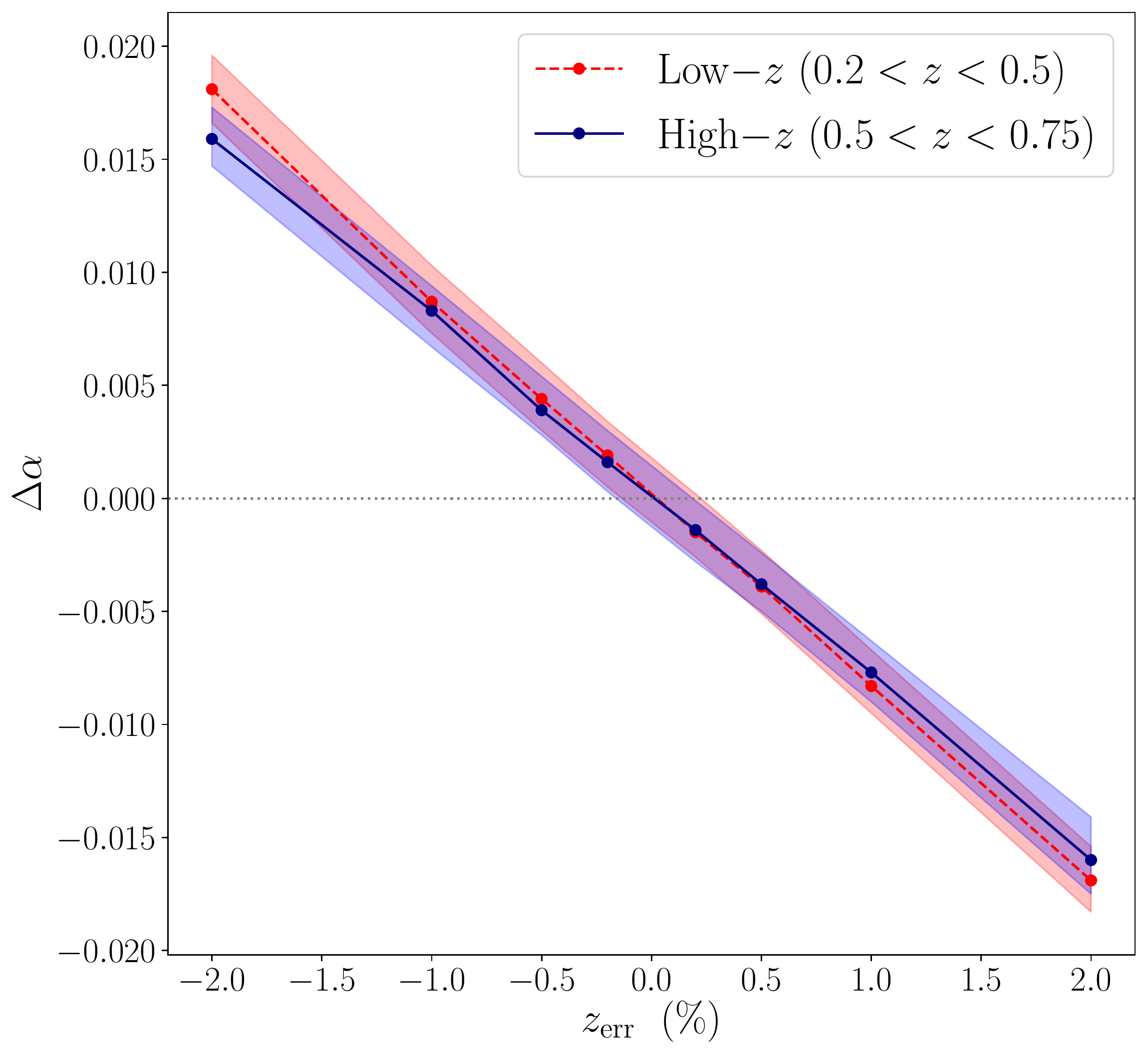}
        \caption[]{Offset in recovered $\alpha$ for a range of injected multiplicative offsets ($z_{\text{err}}$), conducted using mocks that replicate the low and high-$z$ data cuts of BOSS DR12. The trend displayed here is significantly more constrained than the analysis of Figure \ref{fig:BiasedAlpha}, on account of the much larger sample size of the data cuts applied here (when compared to catalogues of fixed $\bar{n}$ used in Figure \ref{fig:BiasedAlpha}). This improved resolution demonstrates a clear linear trend in $\Delta \alpha$ with $z_{\text{err}}$, and highlights the impact of $z_{\text{eff}}$ on the gradient of this relationship (with our low redshift sample demonstrating a slightly higher sensitivity to $z_{\text{err}}$ than our high redshift sample).}
        \label{fig:NGCAlphaOffset}
\end{figure}

We now study the impact of systematic redshift biases on mock catalogues that replicate the BOSS low-$z$ ($0.2<z<0.5, \ z_{\text{eff}} = \Sigma^{n}_i w_i z_i / \Sigma^n_i w_i = 0.38$) and high-$z$ ($0.5<z<0.75, \ z_{\text{eff}} = \Sigma^{n}_i w_i z_i / \Sigma^n_i w_i = 0.61$) redshift bins \citep{Beutler2016}. We construct our baseline sample by applying this low and high-$z$ cut to all 500 Patchy mock samples, corresponding to the populations in Figure \ref{fig:NumDensityComparison}, right. We then inject a series of multiplicative redshift biases, ranging in magnitude from $+2\%$ to $-2\%$. As in the previous section, we fit $\alpha$ for each sample with an injected bias, and compared this to the $\alpha$ of our unbiased reference sample to constrain $\Delta \alpha$ ($\Delta \alpha = \alpha_{\text{bias}} - \alpha_{\text{base}}$). The offset in recovered $\alpha$ arising from these injected redshift biases is shown in Figure \ref{fig:NGCAlphaOffset}.

Broadly, within the explored range of multiplicative offsets, a clear linear relationship between the magnitude of our injected bias and the offset in $\alpha$ emerges. Further, we see a slight difference in the response of our high and low $z$ samples to an injected bias (with our low-$z$ sample demonstrating a slightly higher sensitivity to injected bias), however this effect is minor. This trend is consistent with the model outlined in Appendix A, further re-enforcing the utility of this theoretical model. Due to the improved precision in both low and high-$z$ samples compared to the subsampled fits in Figure \ref{fig:BiasedAlpha}, and the lack of any appreciable dependence between $\Delta \alpha$ and $\bar{n}$, we focus on fitting cosmology to the offsets in $\alpha$ recorded for our low and high-$z$ BOSS-like samples moving forward.

\section{Fitting Cosmological Parameters} \label{sec:cosmo}

\subsection{Single Parameter Fits (BAO Only)}

\begin{figure*}
    \centering
        \includegraphics[width=8.6cm, height=7.5cm]    {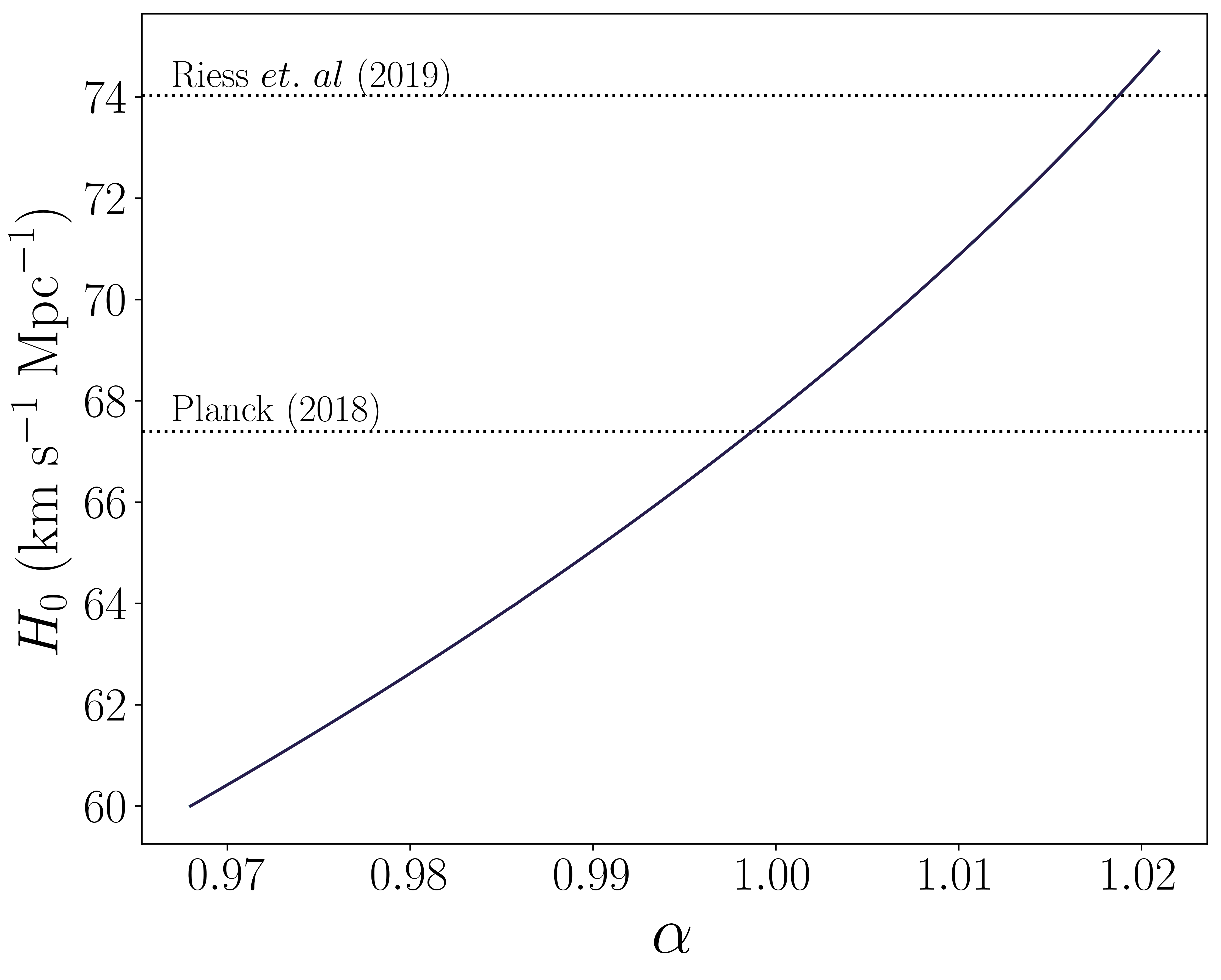}
        \includegraphics[width=8.6cm, height=7.5cm]{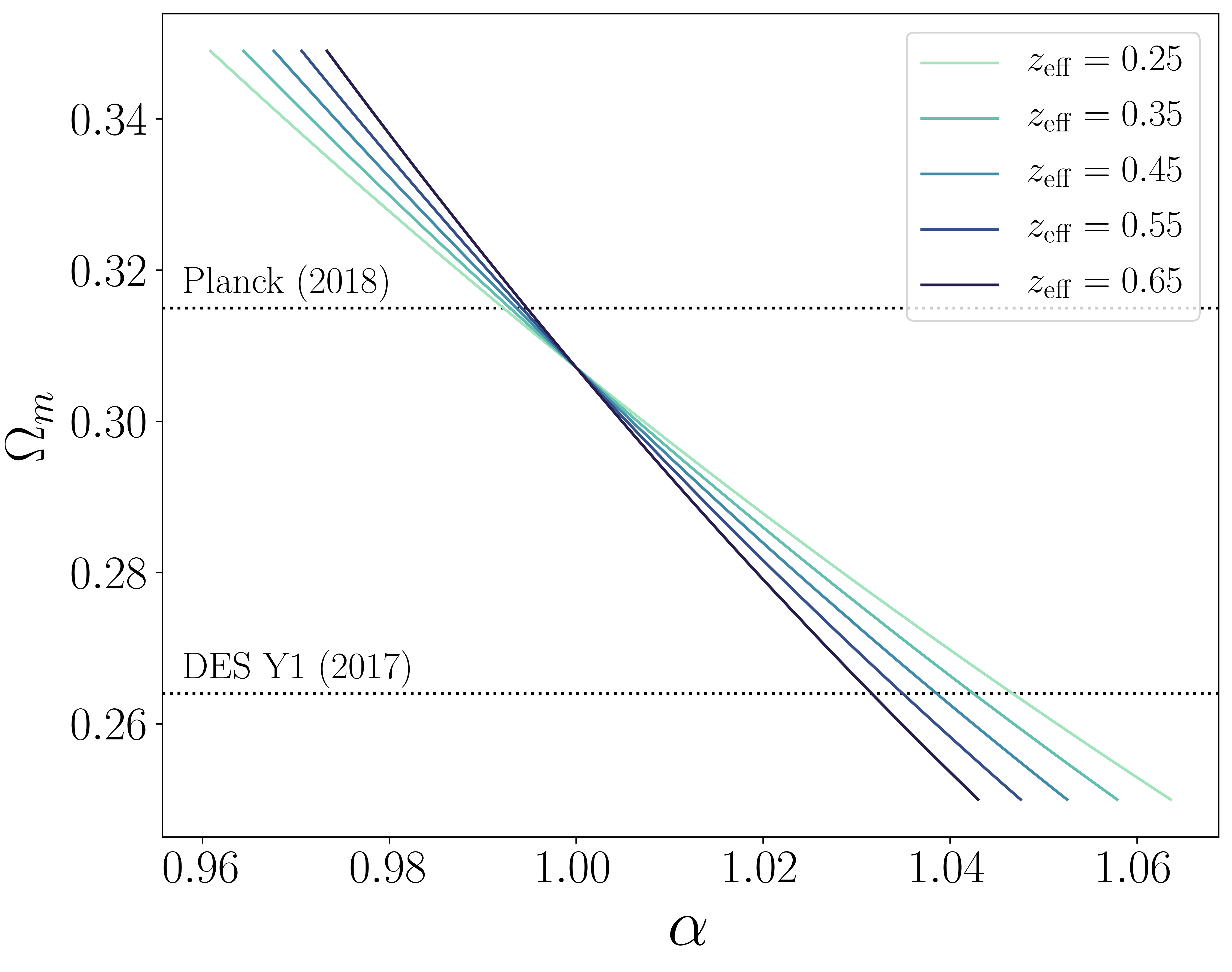}
        \caption[]{Expected $\alpha$ measuring a range of model cosmologies when varying only $H_0$ (left, \kmsmpc), and $\Omega_m$ (right), using the baseline Patchy mock cosmology as our fiducial. Our left plot includes the best fits on $H_0$ derived from \citet{Riess2019} and the \citet{Planck2018}, along with the $\alpha$ one would expect to recover measuring such model cosmologies with respect to our Patchy mock fiducial. Similarly, our right plot includes the best fits on $\Omega_m$ provided by \citet{Planck2018}, along with galaxy clustering and weak lensing constraints from \citet{DESY1}. For a given value of $\alpha$, the corresponding best fit on $H_0$ (when this is the only free parameter) is independent of $z_{\text{eff}}$, yielding overlapping curves. This is not the case when our model $\Omega_m$ is allowed to vary, where the value recovered for a given $\alpha$ is significantly dependent on the $z_{\text{eff}}$ of evaluation. This highlights the need to fold in the updated $z_{\text{eff}}$ of our biased sample when fitting cosmological parameters using $\alpha$, in particular when running simultaneous fits for a range of parameters that include $\Omega_m$.}
        \label{fig:AnalyticAlphavsH0OmegaM}
\end{figure*}

We now turn our attention to exploring how offsets in $\alpha$ arising from some systematic redshift bias can affect cosmological constraints. We begin with an order-of-magnitude analysis using the publicly available Einstein-Boltzmann code \textsc{camb}. We fix the fiducial cosmology of our analysis (used to solve $D_V^{\text{fid}}(z)$ in Equation \ref{eq:alpha}) to the underlying cosmology of the Patchy mock samples. We then solve for the $\alpha$ one would expect to recover when comparing this $D_V^{\text{fid}}$ to a range of model cosmologies (the cosmology used to solve $D_V(z)$ in Equation \ref{eq:alpha}) where $H_0$ ot $\Omega_m$ are allowed to freely vary while all other parameters are fixed. This test, conducted over a range of effective redshifts is given in Figure \ref{fig:AnalyticAlphavsH0OmegaM}.

This test serves to highlight the extraordinary offsets in $\alpha$ required to move between the best-fitting Planck CMB cosmology \cite{Planck2018}, and the constraints provided by works such as \citet{Riess2019} and \cite{DESY1}. Even in a best case scenario when all other parameters are fixed, a $\sim 2\%$ bias in $\alpha$ is required in our BAO fits to replicate a $H_0$ shift of the same magnitude as the Planck-SNe gap. Interpreting this shift as a systematic redshift bias requires the presence of an uncorrected $>2\%$ systematic across all BAO measurements as per Figure $\ref{fig:NGCAlphaOffset}$, far in excess of the $0.2\%-0.4\%$ arising from identified and potentially unresolved spectroscopic systematics. The gap in $\Omega_m$ is even more stark, corresponding to a systematic $3\%-5\%$ shift in $\alpha$ (depending on the $z_{\text{eff}}$ of measurement). It is important to note that the offsets in best fitting cosmology arising from the systematic redshifts biases considered here are almost certainly overestimated, given realistic analyses do not fix all but one cosmological parameter.


\subsection {Simultaneous Fits (BAO + External Probes)}

\begin{figure*}
    \centering
        \includegraphics[width=8.6cm]{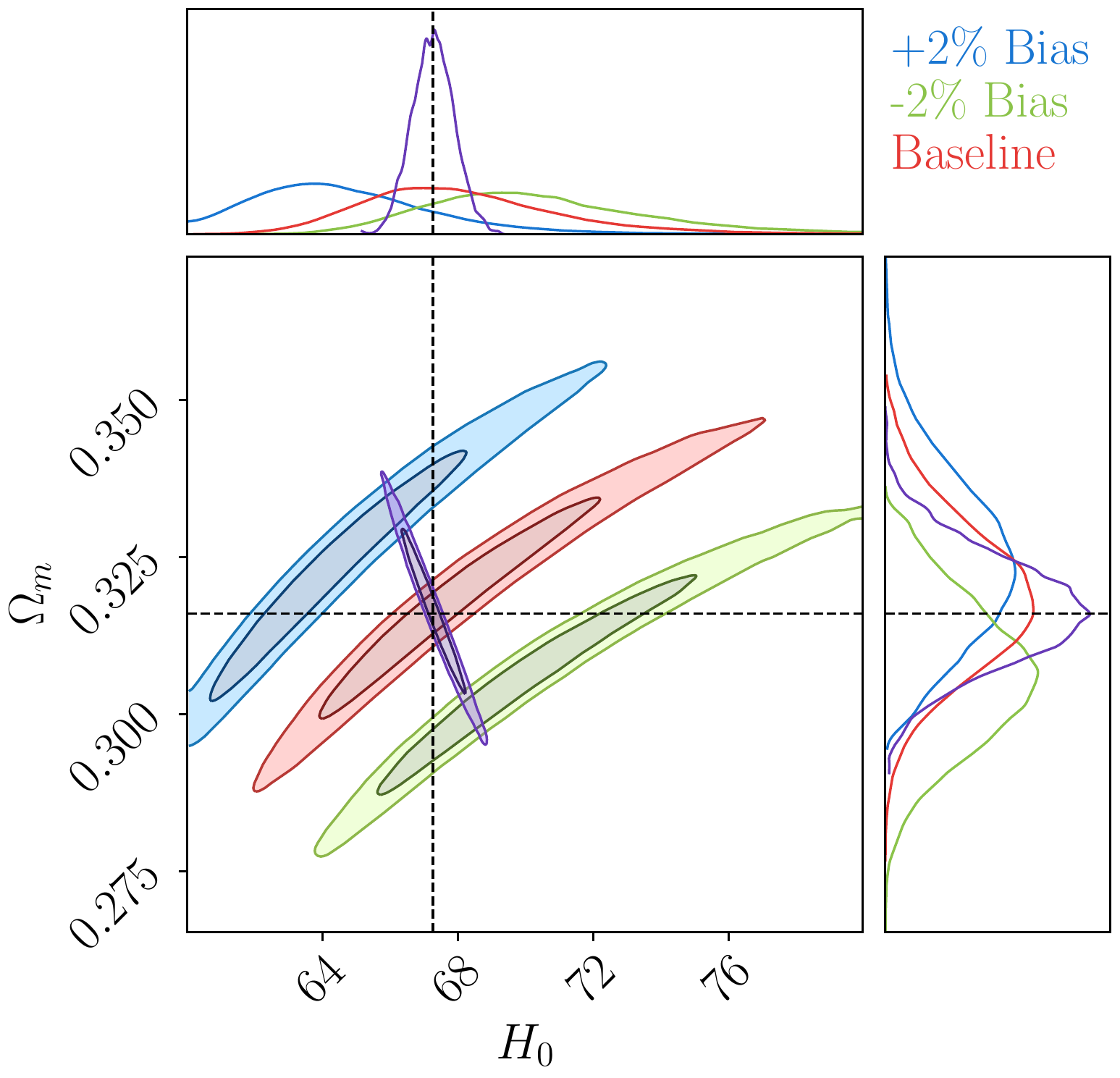}
        \includegraphics[width=8.6cm]{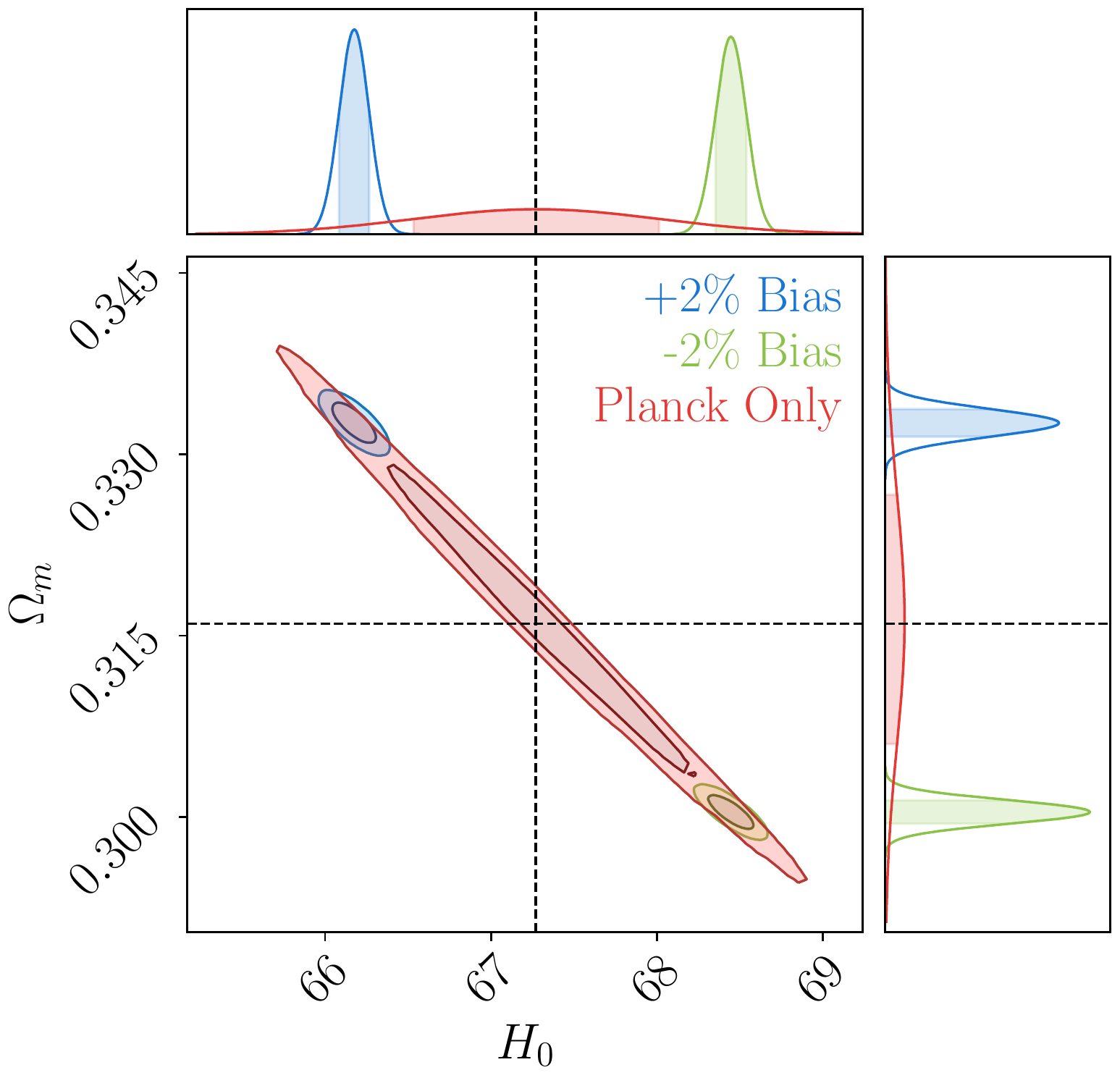}
        \caption[]{Impact of an injected $\pm 2\%$ redshift bias on flat $\Lambda$CDM constraints of $H_0$ (\kmsmpc) and $\Omega_m$ recovered using biased measurements of the BAO feature, and when these measurements are used in conjunction with CMB information from Planck. The left plot corresponds to the shift in BAO only constraints, with the fit provided by Planck data overlapped in Purple. The main effect of this systematic biasing is to shift the BAO contours to higher or lower values of $\Omega_m$. This relationship is consistent with Figures \ref{fig:NGCAlphaOffset} and \ref{fig:AnalyticAlphavsH0OmegaM}, where a systematic upward bias in observed redshift corresponds to a positive shift in $\Omega_m$, and vice-versa. The right plot corresponds to the contours provided by direct importance sampling of the baseline Planck likelihoods, reweighting these likelihoods to include information from our biased BAO sample (as per Equations \ref{eq:reweight} and \ref{eq:LogL}). As the injection of a redshift bias causes the BAO contours to shift along the $\Omega_m$ axis, most of the $H_0$ shift in our importance sampling fit is driven by the orthogonal degeneracy of the Planck CMB contours. Note the magnitude of offsets presented here ($\pm 2\%$) are significantly larger than the combined magnitude of known and potentially unresolved systematics ($~0.2\%-0.4\%$), meaning such effects are unlikely to be present in the literature.}
        \label{fig:2pcOffsetContours}
\end{figure*}

Moving beyond the simple case of single parameter fits, we now focus on how redshift systematics affect simultaneous constraints when information from non-BAO probes are folded into our analysis. In particular, we focus on how redshift systematics (and their corresponding offset in $\alpha$) affect simultaneous fits on $H_0$ and $\Omega_m$ when information from the most recent, publicly available Planck CMB chains are folded into our analysis using importance sampling \citep{Planck2018}. Through this importance sampling technique, we read in a set of points describing the Planck posterior surface, and re-weight each point by the likelihood $\mathcal{L}$ for the biased $\alpha$ constraint, taking into account the original multiplicity weight, $w_{\text{multi}}$, included in the Planck chains,

\begin{equation}
    w_{\text{updated}} = w_{\text{multi}} \times \mathcal{L}.
    \label{eq:reweight}
\end{equation}

The biased $\alpha$ constraint is assumed to come from data with (unknowingly) biased effective redshift $z_{\rm{eff}}$ with a corresponding symmetric, statistical uncertainty $\sigma_{\alpha_{\rm{bias}}}$. We also assume we have conveniently chosen a fiducial cosmological model equal to the maximum \textit{a posteriori} cosmological model of the Planck chain. For each point in the Planck chain with cosmological parameters $\boldsymbol{\theta}$, we can then compute the likelihood as 

\begin{equation}
    \mathrm{ln}(\mathcal{L}) = -\frac{1}{2\sigma_{\alpha_{\rm{bias}}}^{2}} \biggl(\alpha_{\rm{bias}} - \frac{D_{V}(z_{\rm{eff}},\boldsymbol{\theta})r_{s}^{\rm{fid}}(\boldsymbol{\theta})}{D_{V}^{\rm{fid}}(z_{\rm{eff}},\boldsymbol{\theta})r_{s}(\boldsymbol{\theta})}\biggl)^{2}
    \label{eq:LogL}
\end{equation}

From comparison with Eq \ref{eq:alphabiastwo}, we can see that the biased alpha propagates through into biased cosmological constraints because we have assumed our measurement is at the effective redshift $z_{\rm{eff}}$, when in reality it is at a redshift $z_{\rm{eff}}-\Delta z$. In the case of a positive redshift bias, we would spuriously attempt to fit this difference in apparent scale (with our measured scale seemingly smaller than the fiducial evaluated at the true $z_{\rm{eff}}$), resulting in an increase in $\Omega_m$ and a decrease in $H_0$ from our fiducial. Using the offsets in $\alpha$ arising from injected redshift biases in our low and high-$z$ NGC datasets (Figure \ref{fig:NGCAlphaOffset}) as our representative sample, we use this technique to map out the parameter space that simultaneously fits our two biased BAO measurements, and lies within the accepted parameter space of the Planck constraints. 

\subsubsection{BAO+CMB Constraints on Flat $\Lambda$CDM Cosmologies}

\begin{figure}
    \centering
        \includegraphics[width=8.6cm]{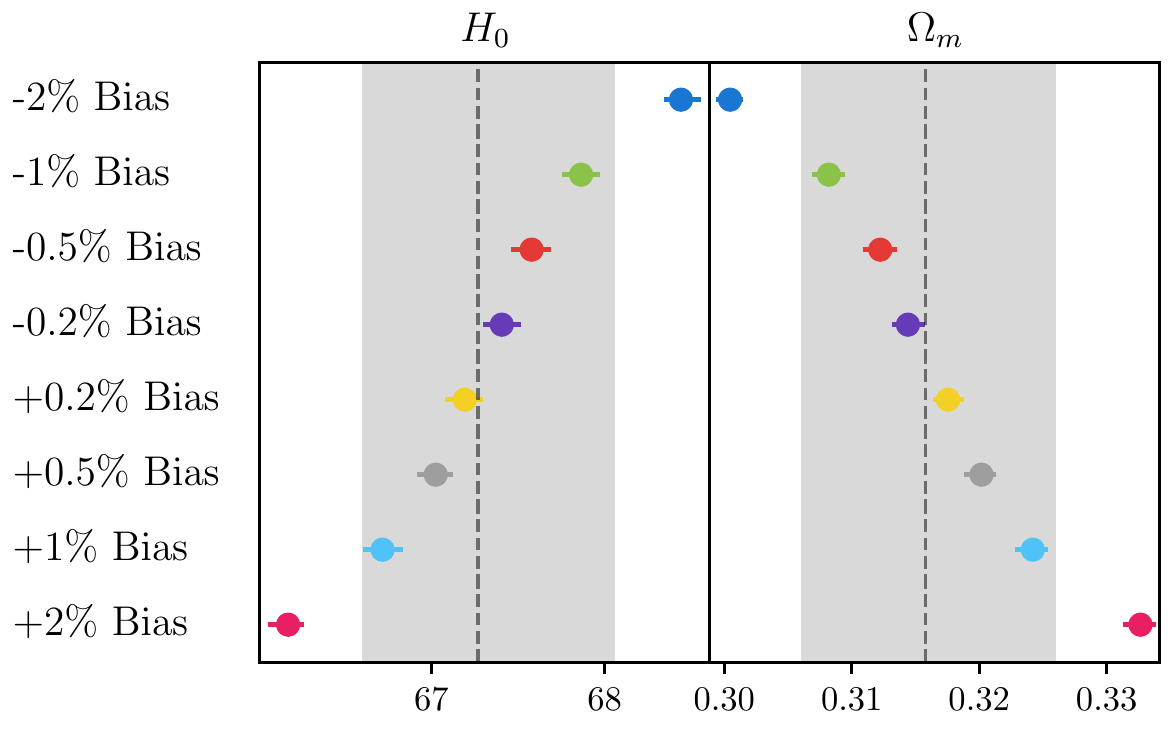}
        \caption[Summary of Parameter Offsets (baseline high-$\ell$ planck power spectra)]{Summary of $H_0$ (\kmsmpc) and $\Omega_m$ fits provided through importance sampling of the baseline Planck CMB spectra for a variety of explored redshift offsets in two BAO measurements. The samples that are biased by the largest explored offset ($\pm 2\%$) prefer values of $H_0$ and $\Omega_m$ that are $\sim2 \sigma$ away from the baseline Planck constraints. Further, within the range of simulated biases, we observe a steady, linear trend in $H_0$ and $\Omega_m$ offsets from the fiducial value as predicted by our simple analytic model. The expected bias from known and unresolved spectroscopic redshift systematics ($\sim 0.2-0.4\%$) is well within the Planck $1\sigma$ error bounds (gray bands).}
        \label{fig:BaselineSummary}
\end{figure}

We begin by resampling the Planck TTTEEE$+$low$\ell$+low$E$ CMB chains for a flat $\Lambda$CDM cosmology, forming our baseline CMB+BAO fit. The distribution we recover from fitting our two largest tested redshift offsets ($\pm2\%$) is given in Figure \ref{fig:2pcOffsetContours}. Interestingly, we find our simultaneous fits using BAO+CMB data recover an offset in $\Omega_m$ that is similar to our BAO only single parameter fits, with a $\pm 2\%$ redshift systematic generating a shift of $\sim 0.02$ in $\Omega_m$ in both cases. In contrast, the simultaneous offset in our best-fitting $H_0$ ($\sim 1 \ \rm{km s}^{-1}\rm{Mpc}^{-1}$) is significantly dampened when compared to our $H_0$, BAO only fit (with a $\sim 5 \ \rm{km s}^{-1}\rm{Mpc}^{-1}$ offset arising from a $2\%$ redshift systematic). By separating the Planck CMB contour from our shifting BAO contours as in Figure \ref{fig:2pcOffsetContours} left, we see that our systematic redshift biases primarily drive variations in the BAO contour along the $\Omega_m$ axis. This makes sense because the change in likelihood due to the redshift bias corresponds to a difference in $D_{v}$ between the biased and true redshifts. $H_{0}$ can be scaled out of this difference, meaning the bias primarily acts on $\Omega_{m}$ for BAO data. When we combine this shift with our Planck data through importance sampling, the posterior distribution remains tightly confined along the diagonal Planck degeneracy direction, driving our simultaneous shift in $H_0$. Repeating this importance sampling technique on the Planck chains for a wide range of redshift biases yields Figure \ref{fig:BaselineSummary}.

We see that the shift in ($H_0, \ \Omega_m$) arising from redshift offsets is uniform throughout the range of explored biases. We see a systematic redshift offset greater than $1\%$ is required to offset our best-fitting ($H_0, \ \Omega_m$) $1 \sigma$ away from the best-fitting baseline Planck constraints. At the level of known potential sources of systematic redshift bias ($\sim 0.2 \%$ at $z_{\text{eff}} = 0.61$ from Figure \ref{fig:RedshiftErrorpc}), we find the bias on our constraints is negligible; on the order of $\sim 0.1 \rm{km s}^{-1}\rm{Mpc}^{-1}$ in $H_0$, and $\sim 2 \times 10^{-3}$ in $\Omega_m$. This suggests cosmological probes that combine BAO constraints with measurements of the CMB are remarkably robust to observational-level redshift biases, requiring injected offsets across multiple measurements that are an order of magnitude larger than known and potentially unresolved effects to significantly bias measurements of $H_0$ and $\Omega_m$.

\subsubsection{BAO+CMB+SNe Ia Constraints on Flat $w$CDM Cosmologies}

Next, we looked at how the inclusion of biased BAO information affects the cosmological constraints recovered from models where the dark energy equation of state ($w$) is allowed to vary. In $w$CDM cosmological models, measurements of the CMB alone provide weak constraints on $H_0$, and require additional information from low redshift probes such as BAO. As a result, it is worthwhile exploring whether this reliance makes flat $w$CDM constraints more sensitive to redshift biases in measurements of the BAO feature than the standard flat $\Lambda$CDM model, or whether the added model flexibility in general acts to protect against such biases.

\begin{figure}
    \centering
        \includegraphics[width=8.6cm]{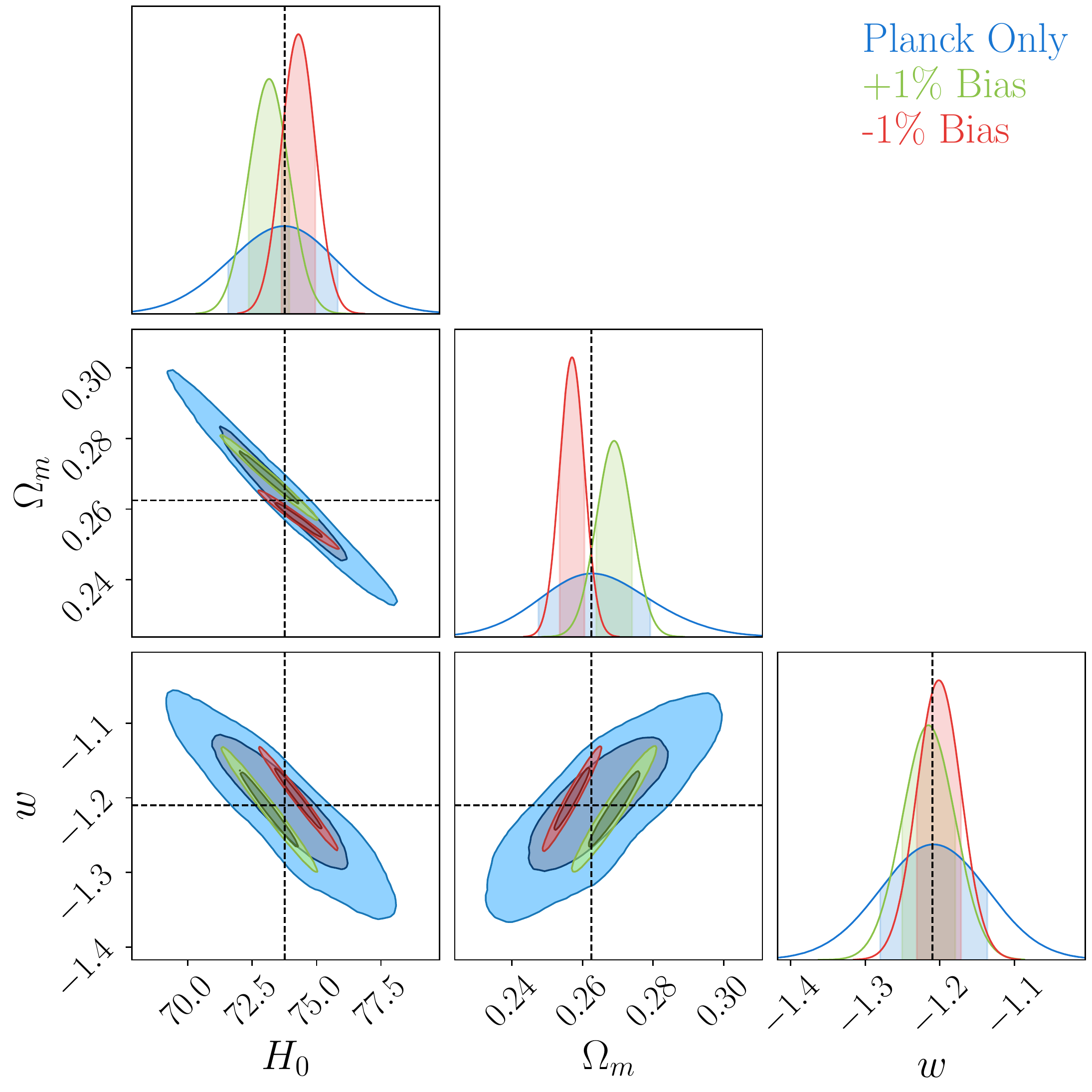}
        \caption[]{Impact of an injected $\pm 1\%$ redshift bias on constraints of $H_0$ (\kmsmpc), $\Omega_m$, and $w$, through importance sampling of Planck $w$CDM chains with a Supernovae-based prior from \citet{Riess2018}. When compared to the flat $\Lambda$CDM fits in Figure \ref{fig:2pcOffsetContours}, we see the injection of a systematic upward redshift bias in BAO measurements results in a positive shift along the $\Omega_m$ axis, and vice-versa. We also see fits to our BAO data within flat $w$CDM models are significantly less constrained than our flat $\Lambda$CDM models. While $w$ as a free parameter is not significantly biased by the presence of redshift systematics in BAO measurements, it introduces notable degeneracies in our parameter slices when compared to flat $\Lambda$CDM fits.}
        \label{fig:pm2pcSupernovaContours}
\end{figure}

\begin{figure}
    \centering
        \includegraphics[width=8.6cm]{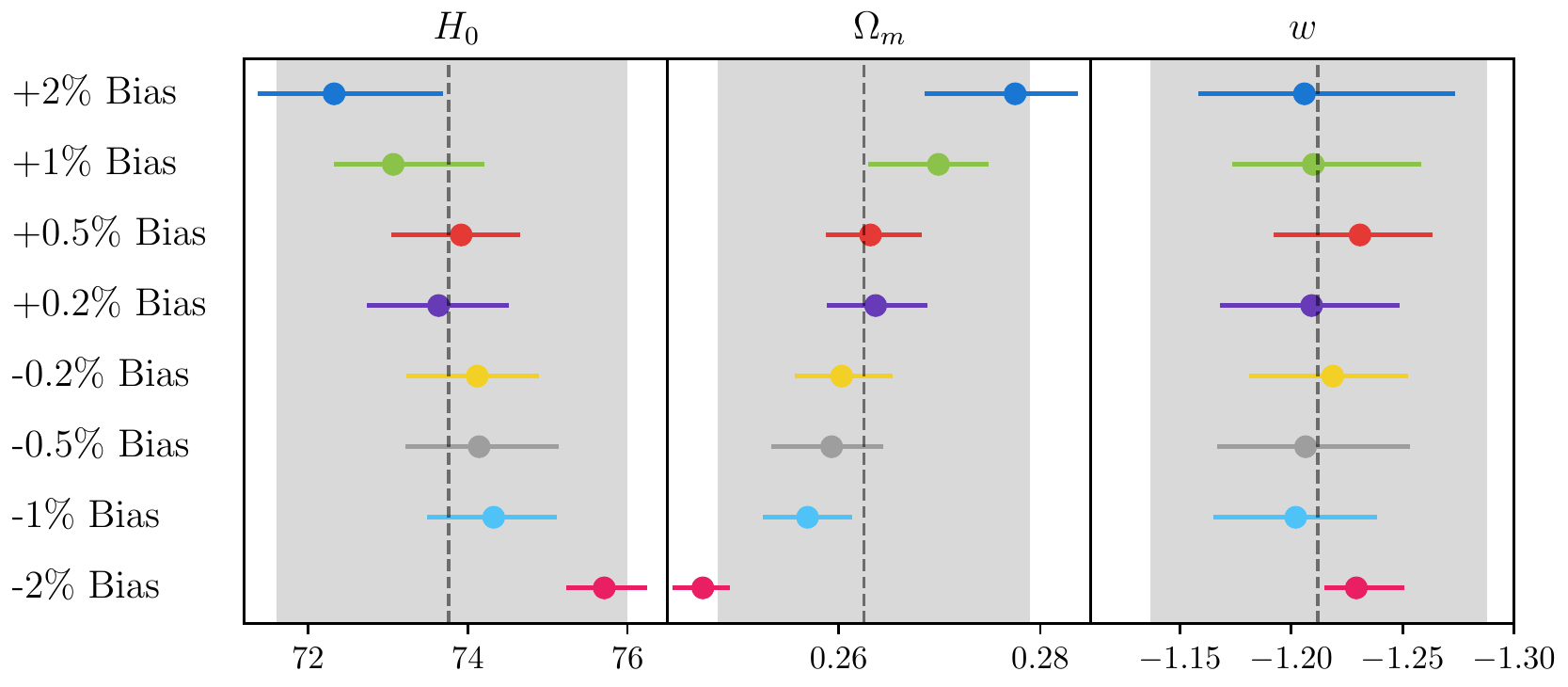}
        \caption[Summary of Parameter Offsets (baseline high-l planck power spectra)]{Summary of $H_0$ (\kmsmpc), $\Omega_m$, and $w$ fits provided through importance sampling of CMB+SNe $w$CDM chains for a range of injected redshift offsets. The added model flexibility provided by allowing $w$ to vary appears to make flat $w$CDM fits less sensitive to small scale ($<0.5\%$) redshift systematics. Beyond this threshold however, flat $w$CDM constraints demonstrate an increased sensitivity to the presence of redshift systematics, particularly in our highest magnitude fits. It is important to note that while flat $w$CDM model fits appear to be sensitive to high-magnitude systematics, their weaker constraining power reduces the \textit{significance} of these offsets compared to the fits of Figure \ref{fig:BaselineSummary}.}
        \label{fig:FullwCDMOffsets}
\end{figure}

We applied our importance sampling technique to Planck TTTEEE$+$low$\ell +$low$E$ CMB chains with a free dark energy equation of state parameter, and a prior provided by supernovae constraints from \citet{Riess2018}. The impact of BAO measurements at $z=0.38$ and $z=0.61$ with a $\pm 1\%$ redshift bias on the cosmological constraints recovered through importance sampling of these Planck $w$CDM + Supernova chains is given in Figure \ref{fig:pm2pcSupernovaContours}, and the shifts recovered for the full range of injected offsets is given in Figure \ref{fig:FullwCDMOffsets}.

When compared to the flat $\Lambda$CDM fits of Figure \ref{fig:BaselineSummary}, we find the increased flexibility provided by allowing $w$ to vary absorbs the effect of low magnitude ($>0.5\%$) BAO systematics. Beyond this low magnitude regime, flat $w$CDM fits exhibit a greater sensitivity to the offsets in $\alpha$ that arise from redshift systematics in measurements of the BAO feature. In particular, when fitting BAO data with a $2\%$ redshift bias, we find the offset in $H_0$ provided by sampling our $w$CDM chains ($\Delta H_0 \sim 2$\kmsmpc) is approximately double the offset recorded for our corresponding $\Lambda$CDM fits. While our flat $w$CDM fits are more \textit{sensitive} to redshift systematics, their reduced constraining power compared to our flat $\Lambda$CDM fits actually reduces the statistical \textit{significance} of these parameter shifts. This emphasises the robust nature of cosmological constraints that combine measurements of the BAO feature with CMB information, demonstrating that even flat $w$CDM models (with a higher degree of freedom than flat $\Lambda$CDM fits) require offsets that are at least an order of magnitude larger than both known and plausibly unresolved effects to significantly bias cosmological constraints.

\subsubsection{BAO + CMB Constraints on $\Lambda$CDM Cosmologies Allowing Curvature}

Finally, we explored how the inclusion of systematically biased BAO information impacts the cosmological constraints recovered from $\Lambda$CDM models where curvature is allowed to freely vary. As detailed in Equation \ref{eq:DA}, the angular diameter distance ($D_A$) varies as a function of curvature, encapsulated by the curvature density parameter $\Omega_k$. Since the BAO feature encodes useful cosmological information in the standard length parameter $D_V$ constructed from this $D_A$, it is important to test the sensitivity of cosmological constraints derived from these standard distance measures to potential deviations from flatness. In particular, we test whether redshift systematics in measurements of the BAO feature yield amplified parameter offsets in models where $\Omega_k$ is allowed to freely vary, when compared to the corresponding parameter shifts in our flat, $\Lambda$CDM model of Section 6.2.1.

\begin{figure}
    \centering
        \includegraphics[width=8.6cm]{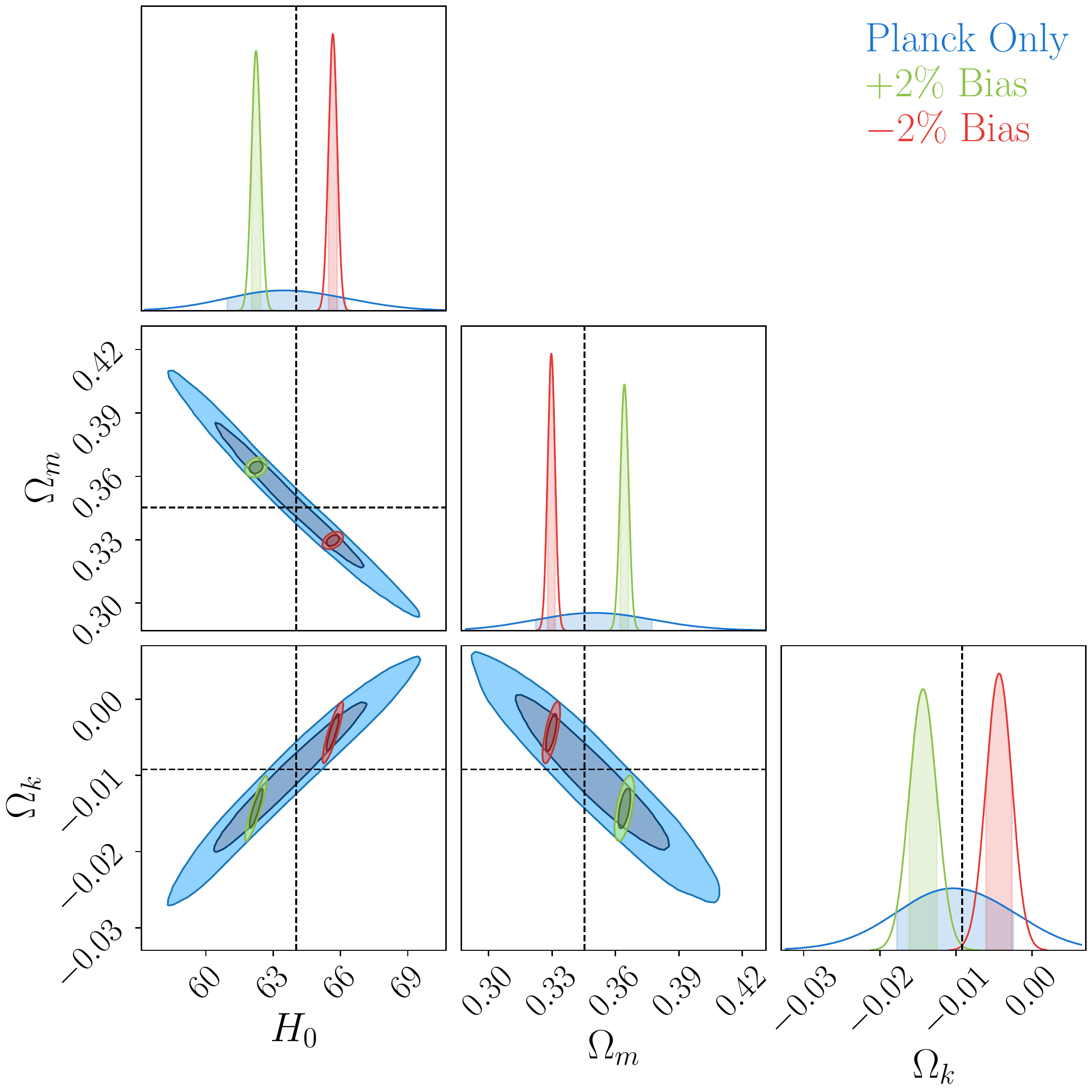}
        \caption[Range of Parameter Offsets $\Lambda$CDM model]{Impact of a $\pm 2\%$ redshift systematic on $H_0$ (\kmsmpc), $\Omega_m$, and $\Omega_k$ constraints within $\Lambda$CDM cosmologies which permit curvature. As in Figure \ref{fig:BaselineSummary} and Figure \ref{fig:FullwCDMOffsets}, we find upward redshift biases work to positively bias constraints on $\Omega_m$ and vice versa. This shift to higher values of $\Omega_m$ corresponds to lower values of $H_0$ and $\Omega_k$ in our full parameter fits. Despite the increased model flexibility provided by removing the condition of flatness, we recover $H_0$ and $\Omega_m$ constraints which remain competitive with our flat $\Lambda$CDM fits, in contrast to the notable degeneracies present in our $w$CDM fits.}
        \label{fig:pm2pcKLCDMOffsets}
\end{figure}

\begin{figure}
    \centering
        \includegraphics[width=8.6cm]{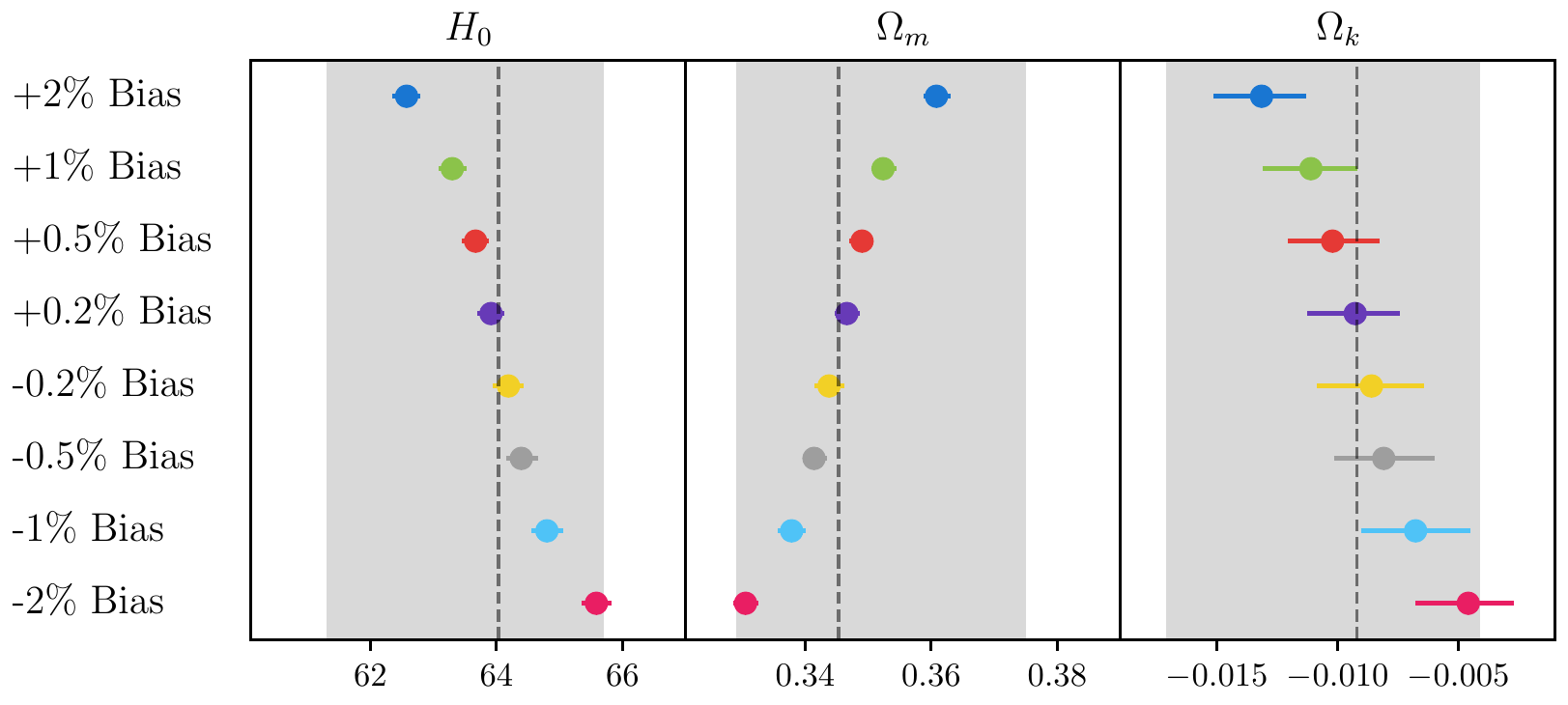}
        \caption[Summary of Parameter Offsets $K\Lambda$CDM model]{The impact of various redshift systematic bias on $\Lambda$CDM models with freely varying curvature. Despite the increased flexibility of $\Lambda$CDM models which permit curvature, we see a significant increase in the magnitude of $H_0, \Omega_m$ shifts in these tests when compared to Figure \ref{fig:BaselineSummary}. Just as in our flat $w$CDM investigation however, the increased model flexibility provided by allowing curvature to vary reduces the statistical significance of these shifts when compared to flat $\Lambda$CDM tests, with the full range of biased fits tested remaining within the $1 \sigma$ error bounds of the original Planck data (grey bands).}
        \label{fig:FullKLCDMOffsets}
\end{figure}

We applied the importance sampling method detailed previously to Planck TTTEEEE$+$low$\ell+$low$E$ chains with a free curvature density parameter $\Omega_k$. Note that the chains tested in this section also include additional information from lensing amplitude reconstruction, as a result of the well reported geometric degeneracy arising in curvature constraints which use CMB power spectrum measurements alone \citep{Planck2018}. First, we highlight the shift in ($\Omega_m$, $H_0$, $\Omega_k$) arising from a $\pm 2\%$ redshift systematic in BAO measurements evaluated at $z_{\text{eff}} = 0.38$ and $z_{\text{eff}} = 0.61$, yielding Figure \ref{fig:pm2pcKLCDMOffsets}. Extending this importance sampling technique over the full range of examined redshift systematics yields the model summary in Figure \ref{fig:FullKLCDMOffsets}.

When compared to the flat $\Lambda$CDM analysis of Figure \ref{fig:BaselineSummary}, we find $\Lambda$CDM models which permit curvature are significantly more sensitive to systematics arising from redshift errors in measurements of the BAO feature. This is well illustrated through the shift in $H_0$ arising from a $\pm 2\%$ redshift systematic, with the $\Lambda$CDM model recovering a $\Delta H_0$ $\sim 2$ \kmsmpc, which is almost twice as large as the corresponding shift in flat $\Lambda$CDM tests. Despite this sensitivity, the increased flexibility provided by removing flatness constraints reduces the statistical significance of these shifts when compared to our flat $\Lambda$CDM tests, with all shifts remaining within the $1\sigma$ region of the Planck constraints. Just as in the case of our flat $w$CDM tests, these results demonstrate that constraints which combine measurements of the CMB and BAO feature are robust to observational redshift systematics across a broad range of extensions to the standard, flat $\Lambda$CDM model.

\section{Discussion and Conclusions}

In this paper, we detail possible sources of low magnitude redshift systematics in galaxy redshift surveys, explore how such systematics could directly impact measurements of the BAO feature, and how these systematics propagate through a variety of cosmological constraints. We find analysis errors such as the use of low-$z$ approximations in heliocentric frame corrections, and potential systematics below the threshold of spectral resolution, could plausibly introduce systematics at an order of $0.2-0.4\%$ at the $z_{\text{eff}}$ of standard BAO surveys. Cosmological constraints that combine measurements of the BAO feature with external probes are incredibly robust to a range of systematic redshift errors, and plausible systematics are very unlikely to bias constraints on parameters such as $H_0$ or $\Omega_m$ now and in the near future.

Using a series of 500 mock galaxy redshift catalogues, we assess the impact of redshift systematics (ranging from $\pm 0.2\%$ up to $\pm 2\%$) on measurements of the BAO feature, and categorise how these bias the dilation parameter $\alpha$. We find a linear relationship exists between the magnitude of an injected bias, and the offset in $\alpha$ derived from these mock catalogues. From first principles, we develop a model to predict the impact of this uniform redshift offset on $\alpha$, and demonstrate this model lies in close agreement with our data. When a uniform redshift offset is introduced into a survey (with respect to some reference catalogue), the underlying physical scale associated with the BAO feature is represented at a different apparent scale. This causes the apparent distance-redshift relationship of a biased sample to predictably deviate from the true distance-redshift relationship of the reference catalogue. 

When BAO measurements are used to make constraints with Planck CMB data in a flat $\Lambda$CDM universe, we find plausible redshift systematics introduce a negligible bias in simultaneous fits on $\Omega_m$ and $H_0$. We determine that observational biases over multiple BAO measurements must be at least an order of magnitude higher than our plausible systematic roof to offset combined constraints beyond $2 \sigma$ from Planck fits. When this analysis is repeated using Planck data with a free equation of state parameter, we find the increased model flexibility makes flat $w$CDM models even more robust to redshift systematics below $<0.5\%$. Beyond this regime, fits to $H_0$ and $\Omega_m$ in flat $w$CDM cosmologies exhibit a greater sensitivity to redshift systematics in BAO measurements than corresponding $\Lambda$CDM constraints. While the magnitude of offsets in ($\Omega_m, \ H_0$) are larger in flat $w$CDM than $\Lambda$CDM models, the uncertainties in our $w$CDM fits also increase significantly. As such, the offsets arising from a $2\%$ systematic redshift bias in measurements of the BAO feature result in parameter shifts within $1\sigma$ of baseline constraints from Planck. When this investigation is repeated for $\Lambda$CDM models with a freely varying curvature, we similarly find fits to $H_0$ and $\Omega_m$ are more sensitive to redshift systematics in measurements of the BAO feature than in flat $\Lambda$CDM models. As in the case of our flat $w$CDM fits however, the increased model flexibility of $\Lambda$CDM models with curvature markedly reduces the statistical significance of such shifts when compared to flat $\Lambda$CDM cosmologies, with the full range of biased fits tested remaining within the $1\sigma$ error bounds of Planck. Constraints which combine measurements of the CMB and BAO feature are remarkably robust to plausible observational redshift systematics across a range of common extensions to the standard, flat $\Lambda$CDM model.

\section*{Acknowledgements}

This research was supported by the Australian Government through the Australian Research Council’s Laureate Fellowship funding scheme (project FL180100168). AG is the recipient of an Australian Government Research Training Program (RTP) Scholarship.

The production of all MultiDark-Patchy mocks for the BOSS Final Data Release has been performed at the BSC Marenostrum supercomputer, the Hydra cluster at the Instituto de Fısica Teorica UAM/CSIC, and NERSC at the Lawrence Berkeley National Laboratory. We acknowledge support from the Spanish MICINNs Consolider-Ingenio 2010 Programme under grant MultiDark CSD2009-00064, MINECO Centro de Excelencia Severo Ochoa Programme under grant SEV- 2012-0249, and grant AYA2014-60641-C2-1-P. The MultiDark-Patchy mocks was an effort led from the IFT UAM-CSIC by F. Prada’s group (C.-H. Chuang, S. Rodriguez-Torres and C. Scoccola) in collaboration with C. Zhao (Tsinghua U.), F.-S. Kitaura (AIP), A. Klypin (NMSU), G. Yepes (UAM), and the BOSS galaxy clustering working group.

\section*{Data Availability}

The underlying mock power spectra and random data sets used in this article were accessed from \url{https://data.sdss.org/sas/dr12/boss/lss/dr12_multidark_patchy_mocks/} and the CMB data used in analysis were accessed from
\url{https://wiki.cosmos.esa.int/planck-legacy-archive/index.php/Cosmological_Parameters}. The code used to perform the analysis in this manuscript will be made available upon reasonable request to the corresponding author. 



\bibliographystyle{mnras}
\bibliography{mnras_template} 




\appendix

\section{Treatment of Redshift Errors}

The use of low-$z$ approximations in converting heliocentric rest frame redshifts to CMB rest frame redshifts introduces significant error. We define this offset ($\Delta z$) as the difference between our naive, low-$z$ approximation ($z_{\rm{naive}}$) and the full conversion ($z_{\rm{true}}$)

\begin{equation}
    \Delta z = z_{\rm{true}} - z_{\rm{naive}} ,
\end{equation}

\begin{equation}
    \Delta z = \left((1 + z_{\rm{obs}})(1 + z^{\rm{pec}}_{\rm{sun}})-1\right) - (z_{\rm{obs}} + z^{\rm{pec}}_{\rm{sun}}).
\end{equation}

Expansion and cancellation of these terms yields,

\begin{equation}
    \Delta z = z_{\rm{obs}} z^{\rm{sun}}_{\rm{pec}}.
\end{equation}

The use of low-$z$ approximations in rest frame corrections offset our recovered redshifts by a product of our observed redshift and the redshift due to our peculiar motion. 
This helps motivate our use of a multiplicative redshift bias in our analysis, but we also consider additive redshift bias to mimic other types of possible systematic bias.







\section{Analytic Estimate of $\alpha_{\text{bias}}$}

Here, we provide an analytic estimate for the expected value of $\Delta \alpha$ for a given galaxy redshift sample with an injected redshift bias ($z_{\text{bias}} = z + \Delta z$), used in Figure \ref{fig:BiasedAlpha}. We begin with the standard form of $\alpha$,

\begin{equation}
    \alpha_{\rm{base}} = \dfrac{D_v(z) r_s^{\rm{fid}}}{D_v^{\rm{fid}}(z) r_s}. \label{eq:alpha_app}
\end{equation}

When some uniform redshift bias is applied our assumed redshift-distance relationship also becomes biased, while the true redshift-distance relationship of the Universe remains the same (the Universe does not care that our measurements of it may be biased). Equation \ref{eq:alpha_app} is hence modified and becomes

\begin{equation}
    \alpha_{\text{bias}} = \dfrac{D_V(z) r_s^{\rm{fid}}}{D_V^{\rm{fid}}(z + \Delta z) r_s}. \label{eq:AlphaBias}
\end{equation}

Expanding our spherically averaged scales $D_v (z)$ and $D_v(z + \Delta z)$,

\begin{equation}
    D_V(z) = \left(\frac{cz D_M(z)^2}{H(z)}\right)^{1/3}
\end{equation}

\begin{equation}
    D_V(z + \Delta z) = \left(\frac{c(z + \Delta z) D_M(z + \Delta z)^2}{H(z + \Delta z)}\right)^{1/3}.
\end{equation}

We approximate $H(z + \Delta z)$ for small $\Delta z$ using a $1^{\text{st}}$ order Taylor expansion

\begin{align}
    H(z + \Delta z) &= H_{0} \left(\Omega_m (1+z+\Delta z)^3 + \Omega_{\Lambda}\right)^{1/2} \\
    &\approx H(z) + \frac{3\Omega_{m}H^{2}_{0}(1+z)^{2}}{2H(z)} \Delta z.
    \label{eq:HzBias}
\end{align}

We then write $D_M (z + \Delta z)$ as,

\begin{equation}
    D_M(z + \Delta z) = c \int^{z}_{0} \frac{dz'}{H(z')} + c \int^{z + \Delta z}_{z} \frac{dz'}{H(z')}.
\end{equation}

which, using the trapezoidal rule, approximates to

\begin{equation}
    D_M(z + \Delta z) \approx D_M(z) + \frac{c \Delta z}{2} \left(\frac{1}{H(z)} + \frac{1}{H(z + \Delta z)}\right).
\end{equation}

Expanding $\frac{1}{H(z + \Delta z)}$ to first order as above and removing small terms ($\propto \Delta z ^2$) yields,

\begin{equation}
    D_M(z + \Delta z) \approx D_M(z) + \frac{c \Delta z}{H(z)}. \label{eq:DmBias}
\end{equation}

Finally, substitution of Equations \ref{eq:HzBias} and \ref{eq:DmBias} into Equation \ref{eq:AlphaBias} and expanding to $1^{\text{st}}$ order again yields,

\begin{align}
    \alpha_{\rm{bias}} \approx \alpha_{\rm{base}} \left[1 - \Delta z \left(\dfrac{1}{3z} + \dfrac{2c}{3H_{\rm{fid}}(z)D^{\rm{fid}}_m(z)} - \dfrac{\Omega^{\rm{fid}}_m H_{0,{\rm{fid}}}^2 (1+z)^2}{2H^{2}_{\rm{fid}}(z)}\right)\right].
    \label{eq:AnalyticAlphaApplied}
\end{align}

This is the approximation used to generate the analytic estimates of Figure \ref{fig:BiasedAlpha}. We can rewrite this expression as,

\begin{align}
    \Delta \alpha &= \alpha_{\rm{bias}} - \alpha_{\rm{base}} \\ \notag
    &\approx - \Delta z \alpha_{\rm{base}} \left[\dfrac{1}{z} + \dfrac{2}{3z}\left(\dfrac{cz}{3H_{\rm{fid}}(z)D^{\rm{fid}}_m(z)} - 1\right) - \dfrac{\Omega^{\rm{fid}}_m H_{0,{\rm{fid}}}^2 (1+z)^2}{2H^{2}_{\rm{fid}}(z)}\right].
\end{align}

The second two terms in this expression are negligible in the limit of low $z$, allowing us to further approximate this as,

\begin{equation}
    \Delta \alpha \approx \frac{- \Delta z}{z}\alpha_{\rm{base}}. \label{eq:lowzDeltaAlpha}
\end{equation}

The approximations given in Equations \ref{eq:AnalyticAlphaApplied} and \ref{eq:lowzDeltaAlpha} are both good to $20\%-30\%$ accuracy up to $z_{\text{eff}} = 0.61$, over a range of cosmologies from $0.1<\Omega_m<0.5$ and $50$ \kmsmpc $< H_0 < 80$ \kmsmpc.


\bsp	
\label{lastpage}
\end{document}